\documentclass{aa}  

\usepackage{graphicx}
\usepackage{txfonts}
\usepackage{lipsum}
\usepackage{subcaption}         
\usepackage{xcolor}
                                
\usepackage{lscape}             
                                
\usepackage{placeins}           
                                
\begin{document}

   \title{T~CrB: overview of the accretion history, Roche-lobe filling,
                 orbital solution, and radiative modeling}

   \titlerunning{Accretion history and radiative modeling of T CrB}

   \author{U.~Munari\inst{1}
           \and
           F.~Walter\inst{2}
           \and
           N.~Masetti\inst{3,4}
           \and
           P.~Valisa\inst{5}
           \and
           S.~Dallaporta\inst{5}
           \and
           A.~Bergamini\inst{5}
	   \and
           G.~Cherini\inst{5}
           \and
           A.~Frigo\inst{5}
           \and
           A.~Maitan\inst{5}
	   \and
           C.~Marino\inst{5}
           \and
           G.~Mazzacurati\inst{5}
           \and
           S.~Moretti\inst{5}
	   \and
           F.~Tabacco\inst{5}
           \and
           S.~Tomaselli\inst{5}
           \and
           A.~Vagnozzi\inst{5}
           \and
           P.~Ochner\inst{6}
           \and
           I.~Albanese\inst{6}
        }

   \authorrunning{U. Munari et al.}

   \institute{INAF National Institute of Astrophysics, Astronomical Observatory of Padova, 36012 Asiago (VI), Italy\\ \email{ulisse.munari@inaf.it}
	\and Department of Physics and Astronomy, Stony Brook University, Stony Brook, NY 11794-3800, USA
	\and INAF Osservatorio di Astrofisica e Scienza dello Spazio, via Gobetti 101, 40129 Bologna, Italy
	\and
             Departamento de Ciencias F\'{i}sicas, Universidad Andr\'{e}s Bello, Fern\'{a}ndez Concha 700, Las Condes, Santiago, Chile
	\and
             ANS Collaboration, c/o Astronomical Observatory, 36012 Asiago (VI), Italy
	\and
	     Department of Physics and Astronomy, University of Padova, Asiago Astrophysical Observatory, 36012 Asiago (VI), Italy
             }

   \date{Received June YY, 2025}

  \abstract
   {Expectations for an imminent new outburst of the recurrent symbiotic
    nova T~CrB are mounting, initiated by the discovery in 2015 of a new
    enhanced mass-transfer phase (SAP), which is reminiscent of the one
    preceding the last recorded outburst in 1946.}
   {We aim to derive a robust estimate of the most important parameters
    describing the physical nature of T~CrB, trace the accretion history
    onto its white dwarf, and account for the unexpected delay in the
    occurrence of the new outburst: the SAP prior to 1946 was brighter, and
    it was followed by the nova eruption within 6 months from its
    conclusion.  This time the 2015-2023 SAP has been fainter and two years
    past its conclusion no new eruption has yet taken place.}
   {During 2005-2025, a period covering SAP and the preceding quiescence,
    we collected a massive amount of photometric and spectroscopic
    observations at optical wavelengths, that we have analyzed together with
    the abundant ultraviolet observations available in the archive of the {\it
    Swift} satellite.}
   {Guided by the results of the orbital solution and in particular by the
    radiative modeling to which we subjected the whole set of available
    data, we found for T~CrB a binary period of 227.5528 days,
    an inclination of 61$^\circ$, and masses of 1.35~M$_\odot$ and
    0.93~M$_\odot$ for the white dwarf and the M3III companion,
    respectively, making mass transfer dynamically stable.  The red giant
    fills completely its Roche lobe, and at $V_{\rm rot} \sin
    i$=4.75$\pm$0.26 km\, s$^{-1}$ it is rotating much slower that the 16
    km\, s$^{-1}$ co-rotation value.  The $\sim$20$^\circ$ azimuth of the
    hot spot, implied by the hump shaping the optical light curve in
    quiescence, fixes the outer radius of the disk to $\sim$58~R$_\odot$,
    the same as the canonical value expected from disk theory.  In
    quiescence the disk is cold and mostly neutral.  SAP has been caused by
    an inside-out collapse of the disk, during which the mean accretion rate
    onto the WD has been $\sim$28$\times$ larger than in quiescence.  SAP
    ended in late April 2023, but from May 2024 mass-flow has intensively
    resumed at disk inner radii while the collapse wave reached the outer
    portions of the disk; the consequent revamp in mass accretion could fill
    the gap inherited by the fainter 2015-2023 SAP and eventually lead the
    WD accreted shell to ignition.}
   {}

   \keywords{binaries: symbiotic -- novae, cataclysmic variables -- Stars: individual: T~CrB -- Accretion, accretion disks}

   \maketitle
   \nolinenumbers

\section{Introduction}

T~CrB is a famous symbiotic nova, i.e.  an otherwise normal nova (powered by
explosive thermonuclear burning) that erupts within a symbiotic binary
\citep[see][for a recent review on symbiotic novae]{2025CoSka..55c..47M}. 
It had outbursts observed in 1866 and 1946 \citep[see the detailed summary
by][and references therein]{1986syst.book.....K}, and possibly others
recorded in ancient times \citep{2023JHA....54..436S}.

 \begin{table*}
 \caption[]{Our $U$$B$$V$$R$$I$ photometry of T~CrB.  The second columns
 provides the HJD$-$2400000.  The full table is available only
 electronically via CDS, a small portion is shown here for guidance on its
 form and content.}
 \label{tab:UBVRI}
 \begin{tabular}{ccccccccccccc}
 \hline \hline
 Date & HJD & $V$ & err & $U-B$ & err & $B-V$ & err & $V-R$ & err & $V-I$ & err & ID\\
 \hline
 2024-01-10.192 &  60319.692 & 9.897 & 0.008  & 0.431 & 0.006   & 1.446 & 0.014  &       &         &       &        &  1402 \\ 
 2024-01-11.114 &  60320.614 & 9.944 & 0.010  & 0.412 & 0.024   & 1.389 & 0.009  &       &         &       &        &  1402 \\ 
 2024-01-11.139 &  60320.639 & 9.935 & 0.005  &       &         & 1.425 & 0.007  & 1.124 & 0.008   & 2.543 & 0.022  &  0310 \\ 
 2024-01-12.141 &  60321.641 & 9.950 & 0.010  & 0.440 & 0.018   & 1.401 & 0.007  &       &         &       &        &  1402 \\ 
 2024-01-13.110 &  60322.610 & 9.966 & 0.011  & 0.763 & 0.018   & 1.443 & 0.009  &       &         &       &        &  1402 \\ 
 2024-01-13.164 &  60322.664 & 9.943 & 0.006  &       &         & 1.415 & 0.022  & 1.149 & 0.009   & 2.528 & 0.016  &  0310 \\ 
 2024-01-14.107 &  60323.607 & 9.948 & 0.010  & 0.503 & 0.015   & 1.456 & 0.016  &       &         &       &        &  1402 \\ 
 2024-01-14.163 &  60323.663 & 9.922 & 0.007  &       &         & 1.456 & 0.006  & 1.116 & 0.006   & 2.606 & 0.018  &  0310 \\ 
 2024-01-15.170 &  60324.670 & 9.967 & 0.008  & 0.410 & 0.007   & 1.366 & 0.009  &       &         &       &        &  1402 \\ 
 2024-01-16.121 &  60325.621 & 9.986 & 0.010  & 0.603 & 0.011   & 1.411 & 0.016  &       &         &       &        &  1402 \\ 
 \hline
 \end{tabular}
 \end{table*}

After the return to quiescence following the 1946 outburst, T~CrB has lived
a quiet life spent accreting on the white dwarf (WD) at (very) low rates,
with just some weak emission in H$\alpha$ generally visible on the optical
spectra otherwise dominated by the absorption spectrum of the M3III red
giant (RG) companion.  On rare occasions \citep{1990JAVSO..19...28I,
1991MNRAS.253..605A, 1998A&A...339..449H}, a temporary surge in accretion
powered stronger emission in the lines and the Balmer continuum.  As
described by \citet{2016NewA...47....7M}, in 2015 T~CrB drastically changed
its status, when it entered a phase of enhanced mass-transfer
which was reminiscent of a similar event, lasting about eight years, which
preceded the 1946 outburst \citep{1946ApJ...104...75P}. For
reasons of continuity with earlier studies of T~CrB, we will keep calling
"SAP" this phase of enhanced mass-transfer onto the WD which precedes an
outburst, even if its original meaning of 'super-active accretion phase' has
in the meantime evolved. The appearance of SAP lead
\citet{2016NewA...47....7M} to consider a new outburst of T~CrB as probable
for 2025-2026, a view which has been since shared by many others
\citep[e.g.][]{2020ApJ...902L..14L, 2023A&A...680L..18Z,
2023MNRAS.524.3146S, 2023AstL...49..501M}.  This has had the beneficial
results of spurring intense pre-eruption monitoring of T~CrB and the
submission of ready-to-be-triggered proposals to most of the ground and
space observing facilities.  Such an alertness from the community,
the expected $V$$\sim$1.5/2.0 mag peak brightness
\citep{1946IAUC.1038....1K, 2024RNAAS...8..233S}, and the proximity of
T~CrB to us \citep[0.92 kpc according to Gaia DR3
parallax;][]{2016A&A...595A...1G,2023A&A...674A...1G}, with all probability
will results is such a wealth of multi-wavelength information to keep people
busy for long in understanding and modeling them, leading to a significant
leap forward in our understanding of novae in general and their symbiotic
subclass in particular.

The SAP phase of T~CrB ended in 2023 \citep{2023RNAAS...7..145M}, lasting
for about the same eight years as the one that preceded the 1946 eruption. 
In this paper we review, on the basis of a massive and multi-wavelength
observational effort we carried out on T~CrB, the ($i$) global properties of
the 2015-2023 SAP in comparison with the long-term preceding quiescence,
($ii$) the amount of mass-flow through the disk to be subsequently accreted
by the WD, ($iii$) the reformation of a steady disk following its collapse
during SAP, ($iv$) revisit of the orbital solution, mass function, and
constraints on the orbital inclination, and finally ($v$) structure, extent
and evolution of the disk during quiescence and SAP as derived by radiative
modeling.

\section{Observations}

The closing date for inclusion of new observations into the present paper is
May 3rd, 2025.

\subsection{UBVRI photometry}

\begin{figure*}
   \centering
   \includegraphics[width=9.1cm]{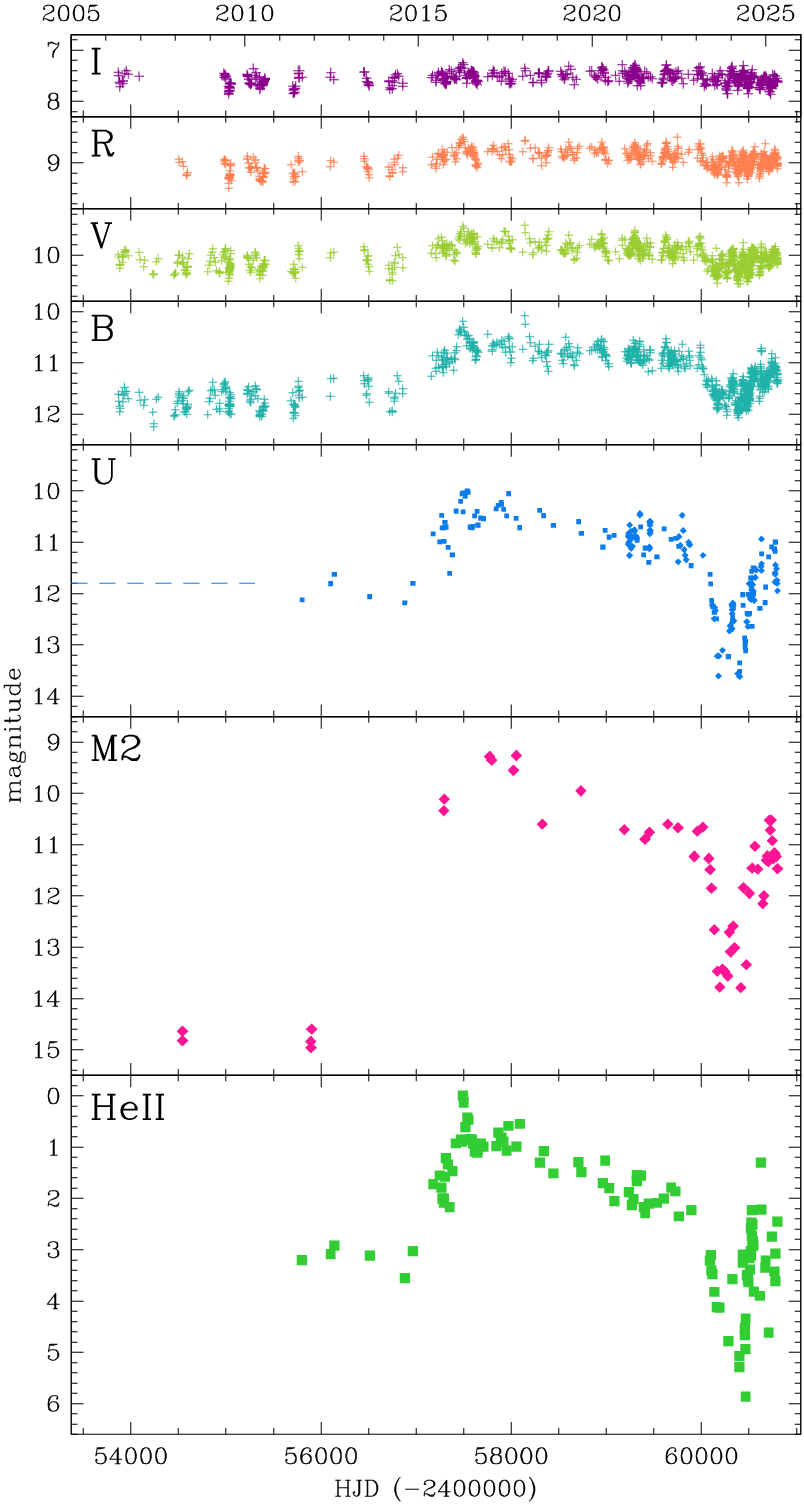}
   \includegraphics[width=8.9cm]{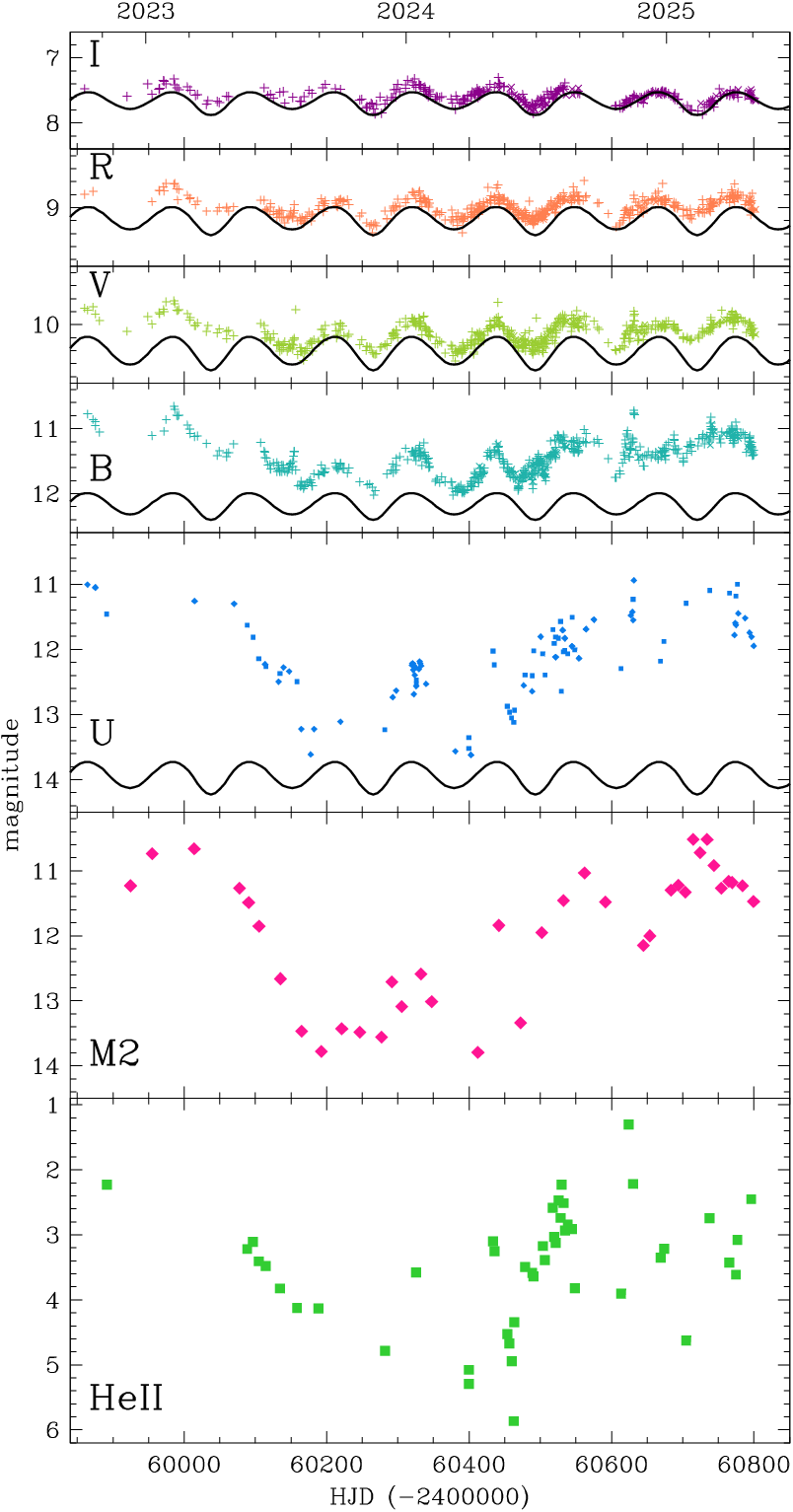}
      \caption{Photometric evolution of T~CrB over the last 20 years, with
      the panels at right zooming on the last four orbital revolutions. 
      $U$$B$$V$$R$$I$ data are from the ANS Collaboration monitoring
      program, M2 is the UVOT band from the {\it Swift} satellite, and the
      flux of the HeII 4686 emission line is expressed in magnitudes
      relative to the peak value ($F_{\rm max}$=4.42$\times$10$^{-12}$
      erg\,cm$^{-2}$\,s$^{-1}$; see sect.~\ref{sect:photometry} for
      details).  The sinusoidal curves at right is the lightcurve in that
      band of the bare, ellipsoidal distorted red giant without contribution
      from the stream, the hot spot, and the accretion disk (see the
      radiative modeling in sect.~\ref{sect:radiative} for details).}
         \label{fig:UBVRI}
   \end{figure*}

$B$$V$$R$$I$ photometry of T~CrB has been regularly obtained since 2005 with
various ANS Collaboration telescopes, while the Asiago Schmidt 67/92cm
telescope (which has an optical train completely transparent to near
ultraviolet wavelengths down to the atmospheric cut-off) has been used
primarily for $U$-band observations.  All telescopes observed T~CrB in all
bands at each visit.  All the photometry has been transformed from the local
instantaneous photometric system to the
\citet{1992AJ....104..340L,2009AJ....137.4186L} standard system via color
equations solved for all frames of each night via the $U$$B$$V$$R$$I$
reference sequence located around T~CrB and calibrated by
\citet{2006A&A...458..339H}.  The same reference photometric sequence and
the same observing/reduction procedures
\citep{2012BaltA..21...13M,2012BaltA..21...22M} have been similarly adopted
at all telescopes over all nights, ensuing a high degree of homogeneity over
the entire 2005-2025 photometric dataset.  The collected photometric data
are listed in Table~\ref{tab:UBVRI}, where the quoted uncertainties are the
total error budget (TEB), which quadratically combine the Poisson error and
the error associated with the transformation to the standard system via the
color equations (usually the dominating term).  Median values for TEB are
0.009 for $V$, 0.012 for ($B-V$), 0.013 for ($V-R$), 0.012 for ($V-I$), and
0.012 for ($U-B$).  The total number of $U$$B$$V$$R$$I$ runs is 712, which
are distributed over 581 different nights.  The corresponding lightcurves
are plotted in Fig.~\ref{fig:UBVRI}, where the large scatter of plotted data
is not due to errors but is instead a manifestation of the orbital
modulation and the large flickering persistently affecting T~CrB.

The number of points in the $U$-band lightcurve of Fig.~\ref{fig:UBVRI} has
been increased by adding the results of integrating the $U$ magnitude on the
low resolution Asiago 1.22m + B\&C spectra (see sect.~\ref{sect:spectra}
below), which are characterized by high S/N, blue wavelength limit around
3200~\AA, and excellent fluxing.  The band transmission profiles of
\citet{1992AJ....104..340L} have been adopted, with the flux zero-points
taken from \citet{1998A&A...333..231B}.  Before computing the $U$ magnitude,
the flux zero-point of the spectra have been scaled so that the computed $B$
magnitude equals that of conventional photometry from Tab.~\ref{tab:UBVRI}. 
To estimate the uncertainty of such a procedure in deriving the $U$ band
magnitude, we have observed over different nights and with the same
instrumentation and data handling procedures about a dozen standard stars of
a red color similar to T~CrB ($B-V$$\geq$0.7), selected from
\citet{2009AJ....137.4186L}.  The $U$ magnitude so computed from the
spectrum did match the tabulated one within 0.1 mag, which is fully adequate
for the aims of the present paper.

\subsection{Spectroscopy}
\label{sect:spectra}

A similarly large effort has been conducted in recording spectra of T~CrB
over 2011-2025 with telescopes located in Asiago, CTIO and Varese.
The spectroscopic data presented in this paper will be made 
public available via CDS.

Low resolution spectroscopy of T~CrB has been obtained over
2011-2025 with the Asiago 1.22m + B\&C telescope, which has an optical train
completely transparent to near ultraviolet wavelengths down to the
atmospheric cut-off.  The detector has been an Andor iDus DU440 CCD camera
(2048$\times$512 pixels, 13.5$\times$13.5~$\mu$m each), which is
characterized by a high near-UV sensitivity. A 300 ln/mm grating blazed
to 5000~\AA\ allowed to cover the 3200-8000
\AA\ range at 2.31 \AA/pix disperion.  The slit was set to a width of
2 arcsec and has always been rotated to the parallactic angle for optimal
flux calibration against the spectrophotometric standards observed each
night.  The slit height of $\sim$8 arcmin allowed a careful sampling and
subtraction of the sky background. 

High resolution spectra of T~CrB were recorded with the Asiago 1.82m
telescope + REOSC Echelle spectrograph.  The 3550-7100~\AA\ interval is
covered in 32 orders without inter-order gaps by an Andor DW436-BV camera
(housing an E2V CCD42-40 AIMO CCD, with 2048x2048 array, and 13.5 $\mu$m
pixel size).  The resolving power is 20,000 for the standard 2-arcsec
slit-width.  The slit height of 22 arcsec allows recording free 8 arcsec on
both sides of the stellar spectrum for a careful definition of the sky
background to subtract.

Echelle spectra of T~CrB were also obtained with the Varese 0.84 m
telescope, equipped with a long-slit mark.III Multi-Mode Spectrograph from
Astrolight Instruments.  The camera is a SBIG ST10XME CCD and the 4250–8850
Å range is covered in 32 orders without inter-order gaps.  A 1$\times$1
binning and the 2.0 arcsec slit width provide a resolving power
$\sim$17,000.

We also observed T~CrB from CTIO in Chile, where it transits low over the
northern horizon, using the CHIRON \citep{2013PASP..125.1336T} fiber-fed
bench-mounted Echelle spectrograph mounted on the 1.5 m telescope operated by
SMARTS.  We used CHIRON in "fiber" mode with 4$\times$4 on-chip binning
yielding a resolution $\lambda$/$\Delta \lambda$$\sim$27,800.
Typical Chiron exposure times were 10~minutes.

All spectra acquired with the Asiago and Varese telescopes have been reduced
in IRAF, with all standard steps involving correction for bias, dark and
flat frames, sky subtraction, wavelength and flux calibration.  Chiron
spectra have been similarly reduced using software written in
IDL\footnote{https://www.astro.sunysb.edu/fwalter/SMARTS/CHIRON/ch\_reduce.pdf}. 
As Chiron is fed by a single 2.5~arcsec diameter fiber, simultaneous sky
subtraction is not possible.  Under clear sky conditions, for a target as
bright as T~CrB, this is an issue only in the vicinity of bright night sky
lines, such as the Na~D lines and [O I] 6300.  Flux calibration is performed
order-by-order using spectra of $\mu$~Col to establish the instrumental
response, and orders are stitched together.  There are five inter-order gaps
longward of 8260 \AA.  Correction to absolute fluxes depends on sky
conditions, and is accomplished by comparison with contemporaneous optical
photometry.

\subsection{Swift satellite ultraviolet observations}\label{sect:Swift}

In order to complement our optical ground-based observations with
ultraviolet (UV) data, we selected a series of pointing performed over
nearly two decades, between 2008 and 2025, with the UltraViolet Optical
Telescope \citep[UVOT;][]{2005SSRv..120...95R} onboard the {\it Neil Gehrels
Swift} satellite \citep{2004ApJ...611.1005G}.  These observations were
downloaded from the ASI-SSDC archive\footnote{{\tt https://www.ssdc.asi.it}}
and were chosen to be as close to simultaneous as possible to specific optical
spectrophotometric epochs we collected during this time range.  UV
observations were acquired in the $UVW1$, $UVM2$ and $UVW2$ filters, with
reference wavelengths 2600, 2246 and 1928 \AA\, respectively \citep[see][for
details]{2008MNRAS.383..627P,2011AIPC.1358..373B}.

All data were reduced within the {\sc ftools} environment
\citep{1995ASPC...77..367B}.  Count rates on Level 2 (i.e., calibrated and
containing astrometric information) UVOT images of T Crb were measured
through aperture photometry within a 5$''$ radius centered on the source
position, whereas the corresponding background was evaluated for each image
using several circular regions in source-free nearby areas.  The UV
magnitudes were then determined using the {\sc uvotsource} task and were
calibrated using the UVOT photometric system described by
\citet{2008MNRAS.383..627P}; the most recent fixings (2020 November)
recommended by the UVOT team were taken into account and a check to reject
small scale sensitivities\footnote{{\tt
https://swift.gsfc.nasa.gov/analysis/ /uvot\_digest/sss\_check.html}} was also
performed.  Furthermore, to the UV images in which the object was saturated
we applied the criterion of \citet{2013MNRAS.436.1684P} to determine the
corresponding magnitudes from the readout streaks produced by the source.

A summary of the UV observations used in this paper, together with the
corresponding results, is presented in Table~\ref{tab:UVOT}.

\section{Reddening}\label{sec:EBV}

The high +48$^\circ$ Galactic latitude implies a low reddening affecting
T~CrB, as supported by published values like $E_{B-V}$=0.15 by
\citet{1982ESASP.176..229C} and $E_{B-V}$=0.07 by
\citet{2022NewA...9701859N}, or the $E_{B-V}$=0.056 upper limit by
\citet{2011ApJ...737..103S}.

\begin{figure}[h!]
   \centering
   \includegraphics[width=\hsize]{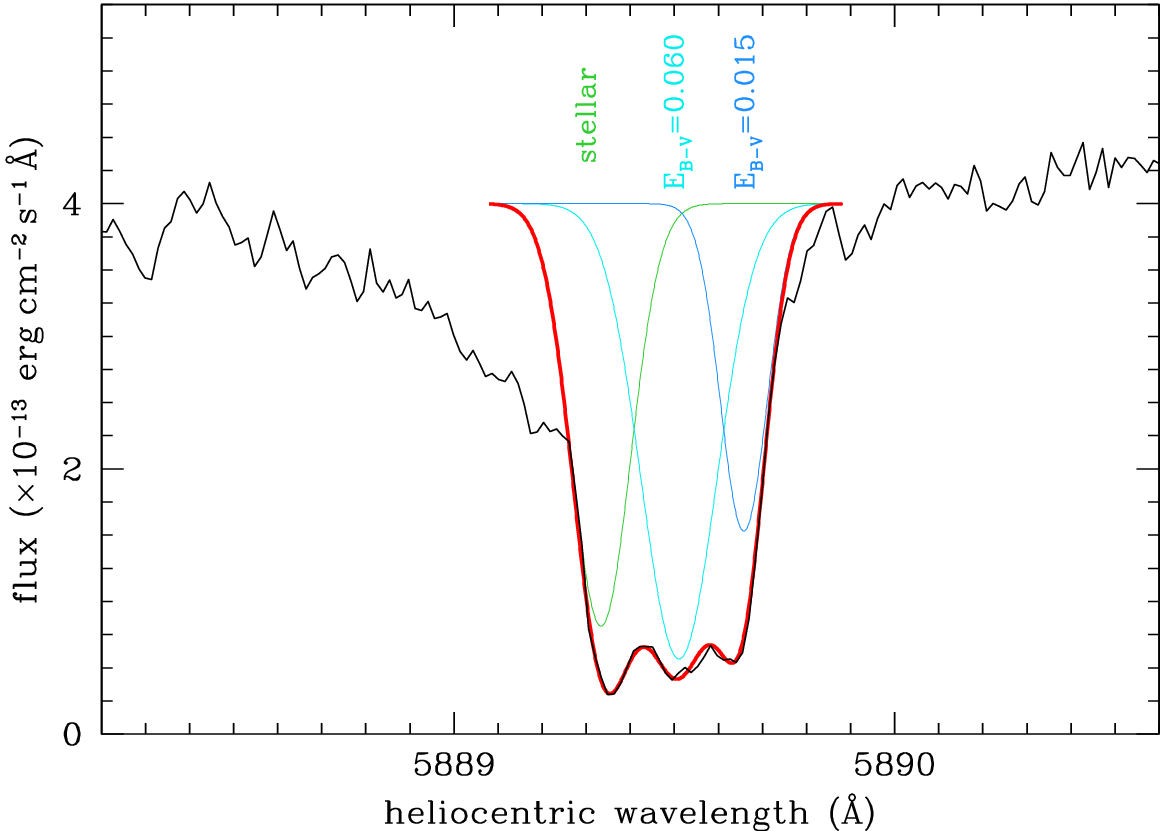}
      \caption{Part of the Chiron spectrum of T~CrB for 2022-08-16, recorded
      in slicer mode for a resolving power of $\sim$78,000.  The NaI line at
      5890 \AA\ is deconvolved into the stellar component (at
      $-$28.5~km\,s$^{-1}$ heliocentric velocity) and additional two of
      interstellar origin (at $-$22.0 and $-$15.0~km\,s$^{-1}$), with
      indication of the corresponding interstellar reddening.}
         \label{fig:NaI}
   \end{figure}

Here we apply an independent method to derive the interstellar reddening. 
All our Chiron Echelle spectra of T~CrB were recorded in fiber mode at a
resolving power of $\sim$27,800, except one.  The spectrum for 2022-08-16
was recorded in slicer mode, boosting the resolving power to $\sim$78,000. 
Such a high value and the large barycentric velocity of T~CrB allows to
disentangle the NaI line profile into the stellar and interstellar
components, as illustrated in Fig.~\ref{fig:NaI} where the profile is fitted
with a combination of three narrow Gaussians (the other NaI line at
5896~\AA\ shows an identical profile).

The bluest of the three fitting Gaussian traces the stellar component at a
measured heliocentric velocity of $-$28.5~km\,s$^{-1}$ well matching the
velocity of the red giant from the orbital solution derived below in
sect.~\ref{sec:Orbit}.  The other two components are of interstellar origin,
and are characterized by heliocentric velocities of $-$22.0 and
$-$15.0~km\,s$^{-1}$, with equivalent widths of 0.1848 and 0.0551 \AA,
respectively.  Adopting the calibration of \citet{1997A&A...318..269M},
these equivalent widths translates into 0.060 and 0.015 reddening values,
for a total $E_{B-V}$=0.075 interstellar extinction affecting T~CrB, that we
will adopt in this paper and which sits in the middle of the range of
reddenings reported in literature.  At such a low reddening, no other
interstellar feature (atomic KI or diffuse interstellar bands) is strong
enough for a fruitful measurement or even detection.

   \begin{figure}[h!]
   \centering
   \includegraphics[width=\hsize]{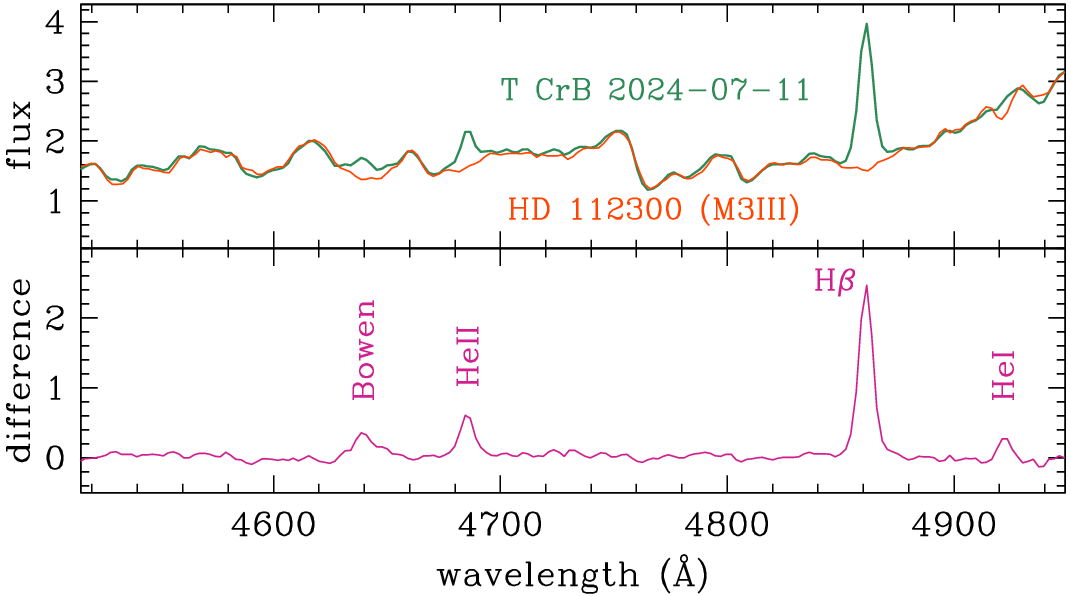}
      \caption{To detect the presence of weak emission lines in the spectra
      of T~CrB, it is necessary to subtract a scaled spectrum of an M3III
      template.  {\em Top panel}: portion of Asiago 1.22m + B\&C + 300 ln/mm
      spectra of T~CrB and the M3III template HD 112300.  {\em Lower panel}:
      the subtracted spectrum reveals the presence of emission lines other
      than H$\beta$.  The ordinates are in
      10$^{-13}$\,erg\,cm$^{-2}$\,s$^{-1}$\,\AA$^{-1}$ on both panels.}
      \label{fig:RGsubtracted}
   \end{figure}

\section{The long-term evolution of T~CrB}
\label{sect:photometry}

The photometric evolution of T~CrB over the last twenty years is plotted in
Fig.~\ref{fig:UBVRI}.  The dashed horizontal line in the $U$-band panel marks the
median quiescence brightness of T~CrB as derived from published observations
\citep[e.g.][]{1986IBVS.2960....1R, 1992A&AS...93..383M,
1997IBVS.4461....1Z, 1998A&A...339..449H, 2004MNRAS.350.1477Z,
2008BaltA..17..293H}.  The panels at right of Fig.~\ref{fig:UBVRI} offer an
expanded view onto the last $\sim$4 orbits of T~CrB, covering the post-SAP
phase.  The continuous curves in each panel show the lightcurve as expected
from the sole ellipsoidal distorted red giant.  These curves have been
computed as part of the radiative modeling performed below in
sect.~\ref{sect:radiative}, to provide a term of comparison against which to
evaluate the contribution of all other components combined (WD, accretion
disk, ionized RG wind, etc.).

To highlight the strict parallelism between accretion-driven photometric
brightness and the spectral appearance, the bottom panel of
Fig.~\ref{fig:UBVRI} presents the flux evolution of the HeII 4686 emission
line as measured on our Asiago 1.22m + B\&C + 300 ln/mm grating spectra. 
For a direct comparison with the photometric magnitudes, in
Fig.~\ref{fig:UBVRI} the flux of the HeII 4686 emission line is also
expressed in magnitudes ($-$2.5$\times$log[Flux/F$_{\rm max}$]), relative to
the peak value $F_{\rm max}$=4.42$\times$10$^{-12}$
erg\,cm$^{-2}$\,s$^{-1}$, which has been reached on our spectrum for April
3, 2016.  The flux of HeII 4686 emission line has been measured on our
Asiago spectra {\em after} the subtraction of a template M3III spectrum (HD
112300) obtained with the same instrumentation and setp-up, as illustrated in
Fig.~\ref{fig:RGsubtracted}, a procedure necessary to provide cleaner and
unbiased results, and which allows detection of HeII at its faintest levels.

   \begin{figure*}[h!]
   \centering
   \includegraphics[width=17cm]{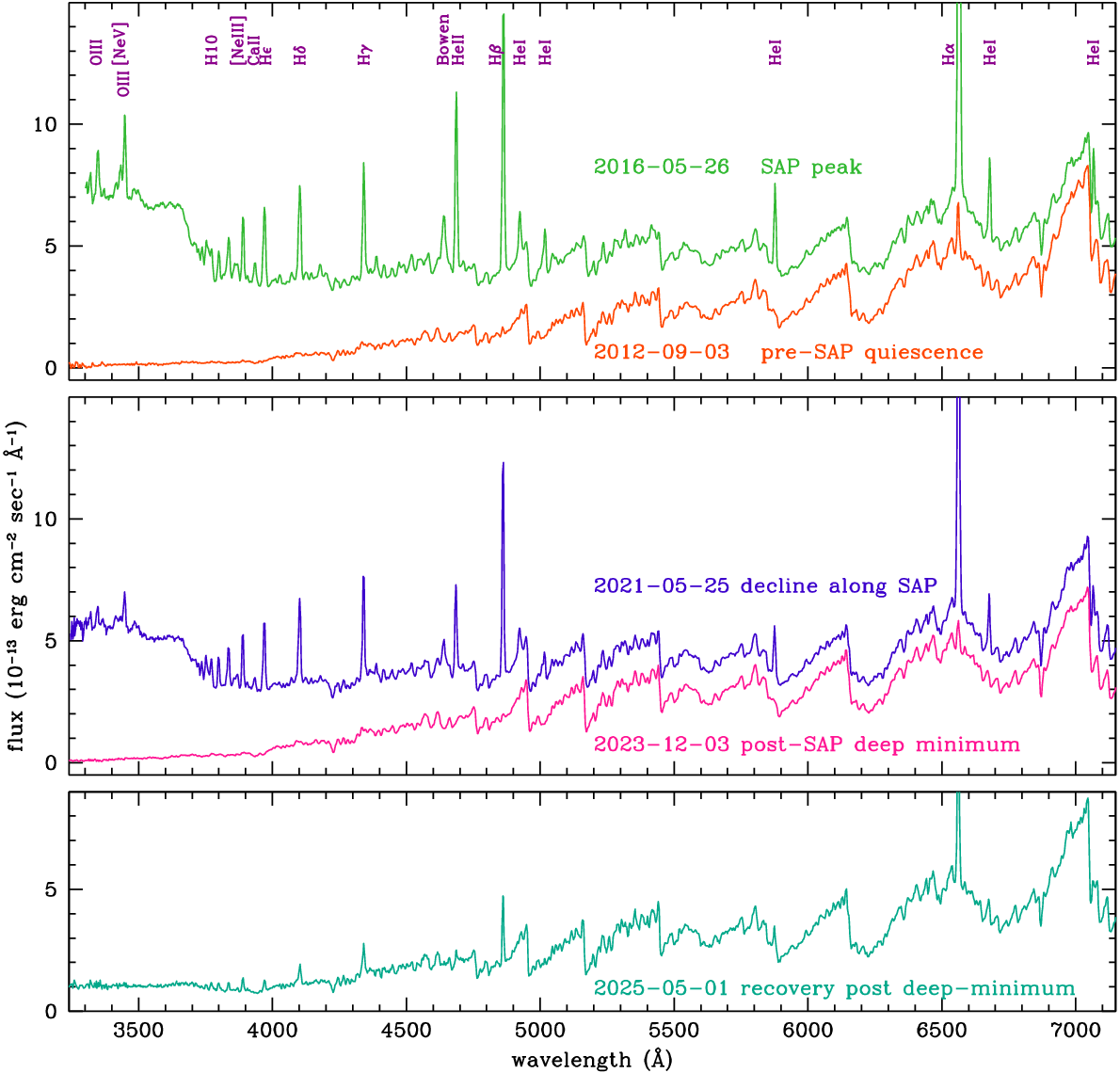}
      \caption{Spectra of T~CrB collected with the Asiago 1.22m telescope
      selected for being representative of the evolutionary phases T~CrB
      has gone through during the last twenty years, as discussed in
      sect.~\ref{sect:photometry}.}
         \label{fig:300tr}
   \end{figure*}

Four phases are clearly distinct in the photometric and spectroscopic
evolution of T~CrB shown in Fig.~\ref{fig:UBVRI}, with corresponding
representative spectra presented in Fig.~\ref{fig:300tr}: 
\begin{itemize}

\item{($1$)} the tail, extending to May 2014, of the long quiescence
following the last outburst in 1946, with the spectrum for 2012-09-03 being
representative of typical appearance during this protracted quiescence,

\item{($2$)} the enhanced mass-transfer phase (SAP), starting in the second
half of 2014, peaking in brightness in May 2016 (cf.  the spectrum for
2016-05-26) and smoothly declining (cf.  the 2021-05-25 spectrum) through
early 2023,

\item{($3$)} a post-SAP deep minimum, with $U$-band brightness and
flux in high-ionization emission lines dropping well below quiescence levels
(cf.  the 2023-12-03 spectrum), and finally

\item{($4$)} a recovery in the accretion flow through the disk and therefore
in brightness (cf.  the spectrum for 2024-11-15), bringing T~CrB back to or
even above the photometric levels of quiescence.

\end{itemize}

\subsection{Phase 1: quiescence prior to SAP}
\label{quiescence}

The period of quiescence of T~CrB following the 1946 outburst and up to the
start of SAP has been already discussed among others by
\citet{2016NewA...47....7M}, and will not be re-investigated here.  The
lightcurve during this phase is dominated by the visibility of the hot spot
and the ellipsoidal distortion of the red giant, which fills its Roche-lobe,
inducing an orbital modulation of increasing amplitude with decreasing
wavelength, reaching $\sim$0.5mag in the $V$ band (see the results of
radiative modeling in sect.~\ref{sect:radiative} below).  The faint status
of the disk around the WD during quiescence allows a clean detection of the
incessant flickering associated with the accretion process \citep[cf.][among
many others]{2001MNRAS.326..553S, 2004MNRAS.350.1477Z, 2010MNRAS.402.2567D}. 
The spectrum for 2012-09-03 in the upper panel of Fig.~\ref{fig:300tr} is
typical of the quiescence period: at a first look it is that of a normal
M3III red giant, the interacting-binary nature betrayed only by limited
emission in H$\alpha$ and a weak flux excess at the shortest wavelengths
caused by the blue continuum originating from the disk and the hot spot.  At
a more careful evaluation, the quiescence spectra reveals the presence of
other and much weaker emission lines (higher Balmer terms, HeI, HeII), but
to get to them it is necessary to subtract the spectrum of a M3III template,
as illustrated in Fig.~\ref{fig:RGsubtracted}.

It is interesting to note that the 2005-2013 quiescence part of the $B$-band
lightcurve of T~CrB in Fig.~\ref{fig:UBVRI} shows a rather slow but
nonetheless distinct rise in brightness, at a rate $\Delta B$=$-$0.040
mag\,yr$^{-1}$, which is probably indicative of an equivalent slow rise in
the accretion flow through the disk.  Similarly slow and low-amplitude
rising and declining trends have been a constant for T~CrB in quiescence
\citep[e.g.][]{2004A&A...415..609S}.

\subsection{Phase 2: the enhanced mass-transfer phase (SAP)}

As first noticed by \citet{2016NewA...47....7M}, T~CrB left quiescence
around May 2014 and begun a steady increase in brightness as a consequence
of a marked and unprecedented increase in the mass flow through the disk and
toward the WD, termed {\em enhanced mass-transfer phase} (SAP) to stress its
uniqueness in the behavior that T~CrB displayed since the 1946 eruption.

The maximum brightness during SAP was reached by T~CrB in April-May 2016,
after a smooth rise lasting two years during which the binary had time to
completed 3 full orbital revolutions.  In comparison with the quiescence
levels in Fig.~\ref{fig:UBVRI}, at SAP peak the brightness of T~CrB increased
by $\Delta U$=3.4 (13.2$\leftrightarrow$9.8), $\Delta B$=1.2
(11.5$\leftrightarrow$10.3), $\Delta V$=0.5 mag (10.1$\leftrightarrow$9.6),
and progressively less for $R$ and $I$ bands; the rise in flux of HeII has
been similar to that of $U$-band, amounting to $\Delta$HeII=3.0 mag.  Soon
after reaching SAP maximum, T~CrB begun a slow and steady descent, at mean 
rates $\Delta$HeII=0.27, $\Delta U$=0.15, and $\Delta B$=0.05 mag\,yr$^{-1}$. 
The SAP phase sharply terminated in early May 2023 (cf.  $U$-band panel at
right in Fig.~\ref{fig:UBVRI}), when T~CrB was still $\Delta U$$\sim$0.5 mag
brighter than average quiescence. Overall, T~CrB spent two years reaching
SAP maximum and seven to decline from and terminate SAP.

The spectrum of T~CrB at SAP maximum is well represented by the observation
for 2016-05-26 in Fig.~\ref{fig:300tr}.  Compared to quiescence (the
2012-09-03 spectrum in the same Fig.~\ref{fig:300tr}), a huge increase in
the continuum flux at blue wavelengths is obvious; as outstanding is the
intensity of emission in the Balmer continuum.  The absolute flux radiated
by emission lines increased by $\sim$15$\times$ compared to quiescence, with
HeII 4686 turning into the third strongest line after H$\alpha$ and
H$\beta$, and HeI and Bowen fluorescence at 4640~\AA\ being rather strong
too.  The difference spectrum (i.e. the SAP spectrum minus that in
quiescence) closely resembles the typical spectrum of a CV in
quiescence, which is dominated by the emission from the accretion disk (to
this end, compare the difference spectrum in Figure~4 of
\citet{2016NewA...47....7M} with the mean CV spectrum of Figure~1 in
\citet{1995A&AS..114..575Z} obtained at the same resolving power). 

The rise to SAP maximum has been characterized by a gradual strengthening of
the emission from the accretion disk (the blue and Balmer continua, the
emission lines), and an increase in the ionization conditions as traced by
the HeII/H$\beta$ and HeII/HeI line ratios.  The decline from SAP maximum
traced back along the same evolutionary pattern followed during the rise:
for ex., the spectrum for 2021-05-25 in Fig.~\ref{fig:300tr}, taken halfway
through the {\em descent} from SAP maximum, is indistinguishable from the spectrum for
2015-10-16 in Figure~4 of \citet{2016NewA...47....7M}, obtained halfway
through the {\em rise} to SAP maximum.

The emission line profiles convey information about the amount of radiating
mass and the kinematical location of the emitting regions and of those in
foreground causing any superimposed absorption components.  With the Asiago
1.82 + Echelle and CTIO 1.5m + Chiron telescopes we densely monitored T~CrB
during the SAP phase, and in Appendix~B we present an atlas of the evolution
of the emission line profiles for a sample of lines, namely H$\alpha$,
H$\beta$, HeII 4686, HeI 5876 (triplet), HeI 6678 (singlet), and [OIII]
5007.  We show the profiles for selected 98 epochs (the same epochs for all
lines), distributed between April 2015 and June 2023, covering the entire
SAP phase.  A detailed analysis and quantitative modeling of such profiles,
with all their fine details and multi-components, responsive to both the
accretion flow and to orbital phase, is well beyond the scopes of the
present paper, and will be tackled elsewhere.

\subsection{Phase 3: the post-SAP deep minimum}\label{deepminimum}

As well illustrated by the $U$ and $M2$ panels in Fig.~\ref{fig:UBVRI}, the
SAP phase ended rather sharply in late April 2023, when T~CrB entered a
steep decline that in four months, by late August, took it close to the
$U$-band brightness of the bare M3III giant, notably {\it well below} the
median $U$-band brightness characterizing the long quiescence since the last
outburst of 1946.  At the time of the late August 2023 minimum, also HeII
dropped below quiescence levels, while $M2$ went down by 3 mag compared to
SAP maximum but remained $\sim$1 mag brighter than in quiescence.

After the rebound in brightness around mid-January 2024, a second drop in
late March 2024 took T~CrB again close to the $U$-band brightness of the
bare M3III.  The two minima (August 2023 and March 2024) were identical in
the $B$$V$$R$$I$ bands and in the {\it Swift} satellite $M2$ ultraviolet
measurments, but they strongly differed in the behavior of HeII, which
dropped almost two magnitudes fainter during the second minimum compared to
the first one.

The spectrum for 2023-12-03 in Fig.~\ref{fig:300tr} well represents average
conditions for T~CrB during the protracted deep minimum phase following the
end of SAP. Comparing with the typical quiescence spectrum for 2012-09-03 in the
same figure, the emission in H$\alpha$ is clearly reduced as well as it is the
veiling from the blue continuum from the disk: compare for example the
depth of blue TiO bands and the visibility of CaI 4227 and CaII H\&K at 3933
and 3967 \AA.  

\subsection{Phase 4: recovery from the deep minimum}\label{recovery}

Following the passage at the second deep minimum in March 2024, T~CrB has
been on a rise in brightness with superimposed a lot of chaotic, short term
and large-amplitude variability, which is particularly evident in the $U$
and HeII lightcurves of Fig.~\ref{fig:UBVRI}, with photometric changes up to
1 magnitude occurring on a matter of a few days and correlating with
revamp/retreat in the intensity of the blue veiling continuum and the emission
lines.

The apparently brightest of such short-living episodes happened on
2024-11-16 when T~CrB peaked at $B$=10.72 and $U$=10.94 as confirmed by
independent observations from three different ANS Collaboration telescopes,
the same three telescopes that recorded T~CrB fainter by 0.6 mag in $B$ and
$U$ both two days before and two days later.  The same flare is well visible
also in the intensity of HeII 4686 (and to a lesser extent in HeI and Balmer
lines too), with the noteworthy feature that in HeII (and marginally in HeI
too) the flare peaks a week earlier than in $U$ and $B$ band photometry. 
Our Echelle spectra obtained with the Varese 0.84m telescopes provide the
integrated HeII 4686 emission line fluxes for November 2024 listed in
Table~\ref{tab:flare}.  Combining with the HeII fluxes from low-res spectra
quoted in \citet{2024ATel16912....1T}, the peak HeII flux was reached on
November 9, with e-folding times of $\leq$1 day for the rise and $\sim$7
days for the decline.

 \begin{table}[t!]
 \caption{Integrated HeII 4686 emission line fluxes (in units of
 10$^{-13}$\,erg\,cm$^{-2}$\,s$^{-1}$) for the flare that T~CrB underwent in
 November 2024.}
 \label{tab:flare}
 \begin{tabular}{rl|rl}
 \hline \hline
 Date (UT)  & HeII flux & Date (UT)  & HeII flux\\
 \hline
  4.72 &   2.1 &  14.73 &  11.1 \\
  7.73 &   5.5 &  15.72 &  12.1 \\
  9.73 &  19.6 &  16.72 &  7.62 \\
 12.72 &  11.5 &  20.72 &  5.73 \\
 13.72 &  12.3 &  22.72 &  4.64 \\
 \hline
 \end{tabular}
 \end{table}

The rise in brightness following the second deep minimum brought T~CrB well
above mean quiescence and close to the level characterizing the latest
stages of SAP, distinctively leveling off during the last few months before
submission of this paper \citep[see also][]{2025arXiv250420592M}.  The
spectrum for 2025-05-01 in Fig.~\ref{fig:300tr} may be considered
representative of average conditions during this leveling off.  Compared
with the typical quiescence spectrum for 2012-09-03 in the same figure,
T~CrB is currently more active, with much stronger emission in Balmer, HeI,
and HeII lines and a Balmer continuum clearly in emission above the
reinforced blue continuum veiling the spectrum of the red giant at the
shortest wavelengths.  After the deep minima of August 2023 and March 2024,
the mass flow through the inner radii of the accretion disk is back at the
levels of the latest stages of SAP and well above that typical for the long
quiescence phase following the 1946 nova eruption.

\subsection{Comparing SAP in 1936-1945 and 2015-2023}

Just prior to the 1946 eruption, T~CrB underwent a SAP phase similar to that
experienced in 2015-2023, a fact leading \citet{2016NewA...47....7M} to be
the first to speculate about an approaching new eruption, a view later
shared by many \citep[eg.][]{2020ApJ...902L..14L, 2023A&A...680L..18Z,
2023JHA....54..436S}, resulting in an intensified pre-eruption monitoring of
T~CrB and the submission of ready-to-be-triggered proposals to most of the
ground and space observing facilities.

   \begin{figure}[ht!]
   \centering
   \includegraphics[width=\hsize]{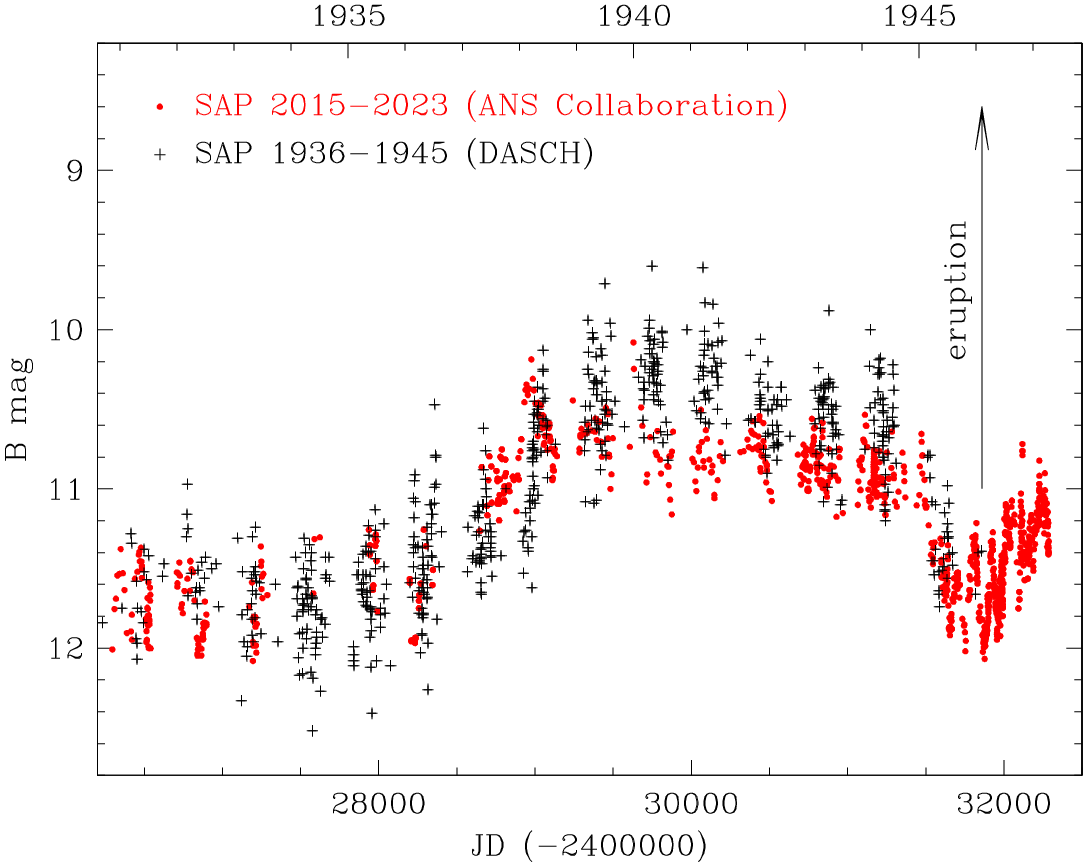}
      \caption{Comparison between the SAP phase that T CrB underwent
      prior to the 1946 outburst and the 2015-2023 recent one.}
         \label{fig:DASCH}
   \end{figure}

The recent and the historic SAP phases (1936-1945 and 2015-2023) are
compared in Fig.~\ref{fig:DASCH}.  The DASCH digitization of Harvard
historical photographic plates \citep{2012IAUS..285...29G} have been
accessed to retrieve the historical data.  The two SAP phases look rather
similar, especially in the overall duration, as well as in the rise- and the
decline-rates.  As discussed below, the SAP event probably corresponds to the
accretion disk flushing toward the WD most of the mass accumulated during
the inter-outburst long interval.  The close similarity of the 1936-1945 and
2015-2023 events suggests that the occurrence of SAP is triggered by the
disk reaching an upper limit to its stable configuration after decades of
steady mass transfer from the Roche-lobe filling RG companion.  There are
however a couple of discrepancies in the two SAP phases:
\begin{itemize}
\item the 1946 nova eruption happened about six months after the SAP
was concluded, while the recent SAP ended more than 2 years ago and T~CrB
has not yet undergone a new outburst;
\item the overall radiated brightness has been larger (by about 40\%) for the SAP
preceding the 1946 nova eruption than for the current one.
\end{itemize} 

The radiated brightness relates to the amount of mass flowing through the
disk, so it is tempting to deduce from the above two points that there has
been yet no new outburst following the end of the 2015-2023 SAP phase
because - compared to 1936-1945 - not enough mass has been so far
transferred from the disk to the WD to reach ignition conditions for a
thermonuclear runaway.  Over the last year, T CrB has however experienced a
recovery to higher-than-quiescence rates in the flow through the disk, which
could make up for the deficit in the mass accumulated on the WD surface and
eventually lead to the much anticipated new eruption.

\section{Radiated luminosity and mass accreted by the WD}

\begin{figure}[th!]
   \centering
   \includegraphics[angle=-90,width=\hsize]{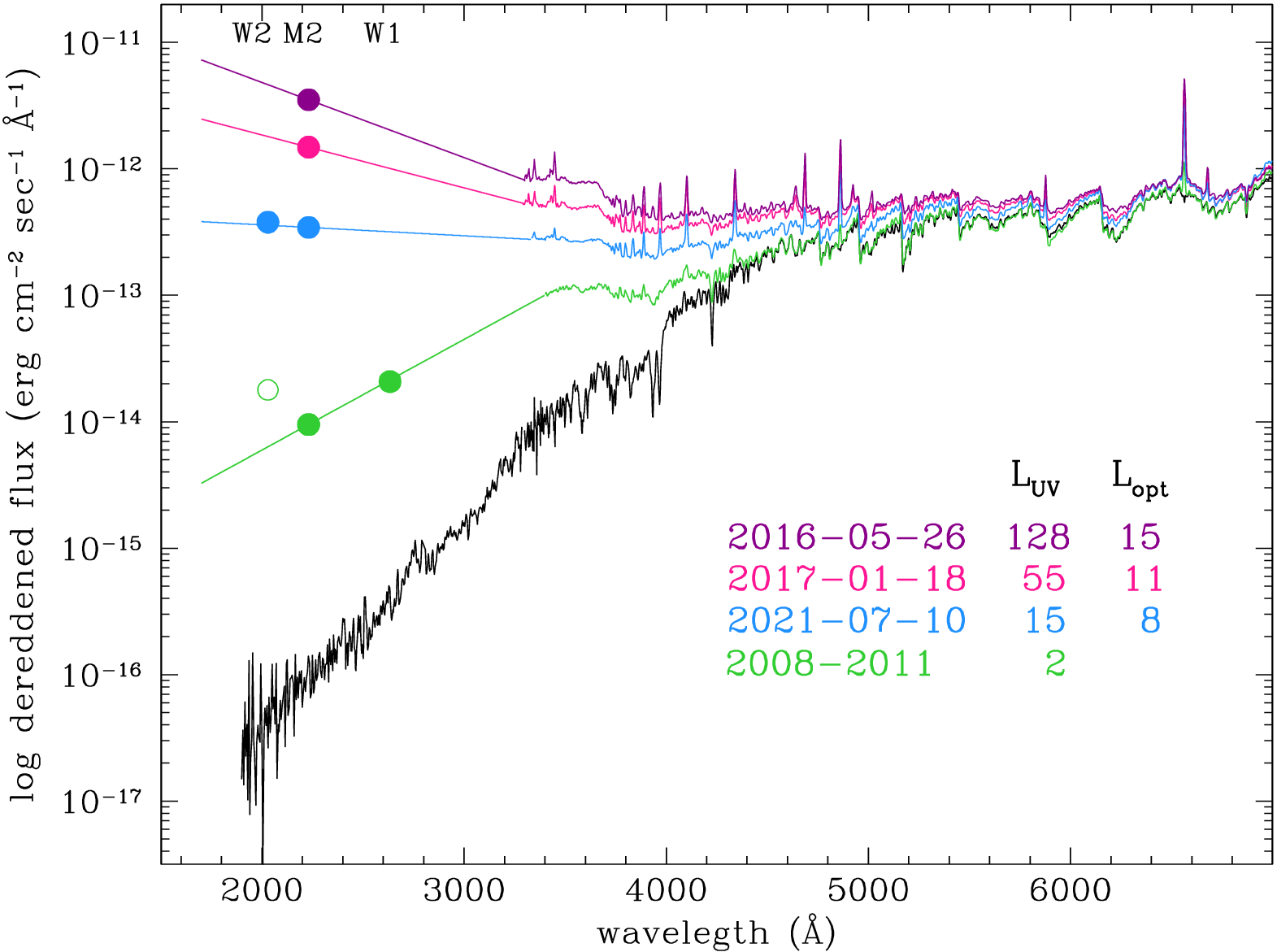}
      \caption{Examples of the spectral energy distribution of T~CrB from
      quiescence (2008-2011) to SAP peak (2016), and for reference that of
      the M3III template HD~112300, plotted in black and scaled to the
      distance of T~CrB.  Quoted $L_{\rm UV}$ and $L_{\rm opt}$ are the
      accretion luminosities (in L$_\odot$) radiated over the 1700-3500 and
      3500-8000 \AA\ intervals, respectively.  Note on the SED for 2008-2011
      the effect of the red-leak affecting the {\it Swift} W2 filter.}
         \label{fig:SED}
   \end{figure}

Combining the {\it Swift} observations in the W2, M2 and W1 ultraviolet
bands with the Asiago 1.22m + B\&C spectra (or $U$$B$$V$$R$$I$ photometry
for the epochs when Asiago spectra are not available), it is possible to
reconstruct fairly well the spectral energy distribution (SED) of T~CrB over
the 1700-8000~\AA\ range, as illustrated in Fig.~\ref{fig:SED}, where the
SED at four sample epochs is compared to that of the template M3III star
HD~112300, constructed from IUE and Asiago spectra, dereddened and
flux-scaled to the Gaia DR3 distance of T~CrB.  By subtracting the SED of
the scaled M3III template to that of any epoch in which a {\it Swift}
observation is available at least in the M2 band, we have obtained the
corresponding SED associated to the accretion process.  Its integration
returns the accretion luminosity radiated over the 1700-8000~\AA\ range; for
illustrative purposes and for a few sample epochs, in Fig.~\ref{fig:SED}
such luminosity is computed separately for the optical (3500-8000 \AA) and
the UV (1700-3500 \AA) parts, to highlight the dominance of the latter.

The {\it Swift} W2 filter is known to be affected by red-leak, which comes
into play when very red objects are observed.  In quiescence, the disk of
T~CrB is faint and most of the flux recorded through the W2 filter (which
nominal ultraviolet transmission starts at 1600 and peaks at 1900~\AA) comes
from the RG through the red leak, making the observations in this filter
useless (cf.  SED for 2008-2011 in Fig.~\ref{fig:SED}).  When however the disk
brightens and the ultraviolet flux increases, as during SAP, the effect of
the red-leak becomes irrelevant and the W2 measurements can be safely
incorporated into building the SED (cf.  SED for 2021-07-10 in
Fig.~\ref{fig:SED}).

The SAP and following phases have been fairly well covered by {\it Swift}
and near-simultaneous optical observations, so that the evolution of the
accretion luminosity can be accurately followed, with the caveat that {\it
Swift} did not observe T~CrB during the broad peak of 2016.  The 2016 has
been however densely covered by spectroscopic observations, and the
integrated flux of the HeII 4686 line can be used to estimate the
corresponding flux at M2 wavelengths.  To this aim we have compared the
integrated flux of the HeII 4686 line and the flux at M2 effective
wavelength for the many SAP epochs in which the two values have been
measured within a 2-day delay, to minimize the disturbance by the short term
variability characterizing T~CrB.  We found a good proportionality of the
two reddening-corrected values in the form:
\begin{equation} 
\log({\rm flux [M2]}) = 0.927\times\lg({\rm flux [HeII]}) - 1.515 
\end{equation} 

In the process we found that the integrated flux of the HeII 4686 emission
line also strongly correlates with the $U$-band brightness in the form
corrected for reddening:
\begin{equation} 
\log({\rm flux [HeII]}) = - 0.5025\times U - 6.229 
\end{equation}
In both relations the flux is expressed in erg\,cm$^{-2}$\,s$^{-1}$.  Eq.(1)
then allows us to reconstruct the SED also during the broad SAP peak of 2016
by using the many measurements available for HeII.

By integrating the SEDs over 2015-2025, we found that during SAP T~CrB
radiated over the interval 1700-8000~\AA\ a total accretion luminosity of:
\begin{equation}
L_{\rm acc}^{\rm SAP}{\rm [UV-opt]} = 6.7\times 10^{43}~~~{\rm erg}
\end{equation}
with
\begin{equation}
M_{\rm acc}^{\rm SAP}{\rm [UV-opt]} = 2.6\times 10^{-7}~~{\rm M}_\odot
\end{equation}
as the corresponding mass accreted by the WD of 1.35~M$_\odot$; 
such a total splits into yearly percentages (\%) of 9.4, 19.5,
18.4, 15.0, 12.4, 7.7, 6.3, 5.5, 2.5, 1.9, and 2.8 going from 2015 to 2025.

The quiescence phase following the 1946 eruption has not been mapped equally
well.  For sake of discussion and considering the self-similar spectroscopic
appearance that T~CrB maintained during the whole quiescence, we may suppose
that the accretion luminosity $L_{\rm acc}$$\sim$2 L$_\odot$ derived in
Fig.~\ref{fig:SED} as the mean value for 2008-2011 apply also to the rest of
the quiescence; under such circumstances the total accretion luminosity 
radiated over the interval 1700-8000~\AA\ during 1947-2014
then becomes:
\begin{equation}
L_{\rm acc}^{\rm quiesc}{\rm [UV-opt]} = 1.6\times 10^{43}~~~{\rm erg}
\end{equation}
with a corresponding accreted mass of
\begin{equation}
M_{\rm acc}^{\rm quiesc}{\rm [UV-opt]} = 6.3\times 10^{-8}~~{\rm M}_\odot
\end{equation}

The sum of $M_{\rm acc}^{\rm SAP}{\rm [UV-opt]}$ and $M_{\rm
acc}^{\rm quiesc}{\rm [UV-opt]}$ leads to $M_{\rm acc}{\rm
[UV-opt]}$=3.2($\pm$1.1)$\times 10^{-7}$ M$_\odot$ as the mass (so far)
accreted in the inter-outburst period.  This value stems from the accretion
luminosity radiated longword of 1700~\AA, so it has clearly to be considered
as a lower limit to the true effective value \citep[see for
ex.][]{2015NewA...36..116S}.  The mass needed to trigger the outburst in
conditions resembling those of T CrB ranges over 0.2$-$2$\times
10^{-6}$~M$_\odot$ according to various theoretical models, as for example
those of \citet{2005ApJ...623..398Y}, \citet{2009ApJ...692..324S}, or
\citet{2025ApJ...982...89S}, much depending on the assumptions - in addition
to differences in the models - on the composition and radial structure of
the underlying WD, how the chemical mixing between the accreted shell and
the underlying WD actually takes place, the detailed chemical and isotopic
composition of the matter accreted from the companion, and so forth.  What
is worth noticing here is the satisfactory agreement between the accreted
mass we derived above from available observations and that required in
theoretical models to ignite an explosive outburst for the conditions
prevailing in T CrB.

\section{Orbital solution}\label{sec:Orbit}

The radial velocity (RV) of the RG has been measured via cross-correlation
on the high-resolution spectra of T~CrB that we collected with the Asiago
1.82m + Echelle and SMARTS 1.5m + CHIRON telescopes, with
Table~\ref{tab:appendix-RV} reporting the results of our measurements.  The
template for cross-correlation was selected from the synthetic spectral
library of \citet{2005A&A...442.1127M} for the resolving power 20,000, with
atmospheric parameters $T_{\rm eff}$=3500~K, $\log g$=1.5, [Fe/H]=0.0,
[$\alpha$/H]=0.0, $\zeta$=2~km\,s$^{-1}$.  The choice of a synthetic
spectrum nulls the combined effects of telluric absorptions, limited S/N,
and errors in the wavelength calibration intrinsic to any RV standards
observed nightly along with the target.  In addition, a synthetic spectrum
also nulls the effect of the jitter caused by the atmosphere of a cool giant
to be deeply convective, extended, and ultimately too unstable to serve as
an accurate RV standard.  To check for any offset to RVs that the choice of
a synthetic spectrum could introduce, we have selected from the Gaia DR3
catalog eight M2-M4 giants among those listed with the lowest uncertainty on
their RV, and observed them with the Asiago 1.82m + Echelle telescope along
with T~CrB.  Their RVs were derived via cross-correlation against the same
synthetic template spectrum used for T~CrB.  The mean difference with the RV
listed in Gaia DR3 is 0.10$\pm$0.27 km\,s$^{-1}$, confirming the absence of
any systematic offset.

   \begin{figure}[ht!]
        \centering
        \includegraphics[width=\hsize]{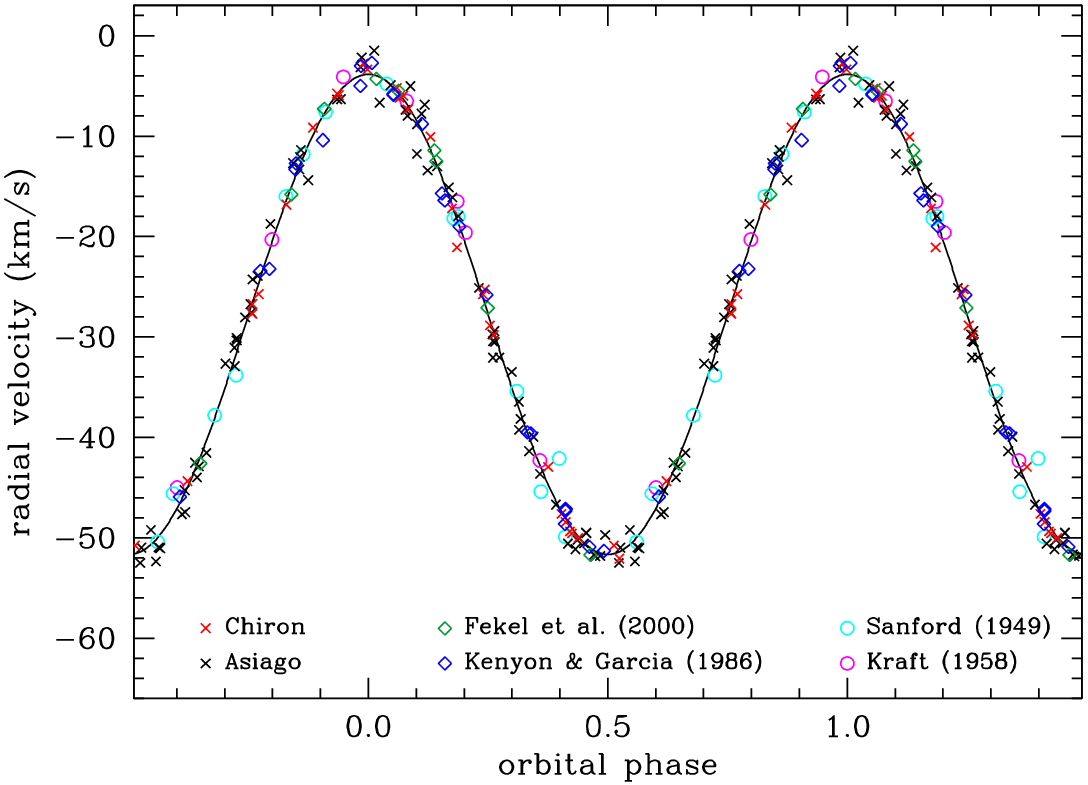}
        \caption{The orbital solution derived in Table~\ref{tab:orbit}
        overplotted to our Asiago and Chiron radial velocities and those of
        \citet{1949ApJ...109...81S}, \citet{1958ApJ...127..625K},
        \citet{1986AJ.....91..125K}, and \citet{2000AJ....119.1375F}.}
        \label{fig:orbit}
    \end{figure}

 \begin{table}[ht!]
 \caption{Spectroscopic orbit of the M3III in T~CrB derived by combining our
        Asiago and Chiron radial velocities with those listed by
        \citet{1949ApJ...109...81S}, \citet{1958ApJ...127..625K},
        \citet{1986AJ.....91..125K}, and \citet{2000AJ....119.1375F}.  $T_0$
        is the time of passage at ascending quadrature (maximum velocity).}
 \label{tab:orbit}
 \begin{tabular}{rll}
 \hline \hline
 Parameter&Value&Error\\
 \hline
 Period (days)                       &    227.5528 & 0.0002  \\
 $T_0$ (HJD)                         &  2459978.37 & 0.08   \\
 Barycentric velocity (km\,s$^{-1}$) &    $-$27.75 & 0.04   \\
 Semi-amplitude (km\,s$^{-1}$)       &       23.90 & 0.05   \\
 Eccentricity                        &           0 &        \\
 Mass function                       &       0.322 & 0.002  \\
 \hline
 \end{tabular}
 \end{table}
  
A spectroscopic orbit for the cool giant in T~CrB has been previously
derived by \citet{1986AJ.....91..125K} and \citet{2000AJ....119.1375F}. 
Combining our RVs with theirs, and including the 1946-47 data
of \citet{1949ApJ...109...81S} and the 1956-57 from
\citet{1958ApJ...127..625K} allow us to cover a time-span of more than 28200
days ($\sim$124 orbits), well constraining the orbital parameters and the
period in particular.  The resulting orbital solution for the RG in T~CrB is
listed in Table~\ref{tab:orbit} and overplotted to the data in
Fig.~\ref{fig:orbit}.  A null eccentricity is clearly supported by the data,
in agreement with previous studies, and we fixed it to 0 in deriving the
orbit.  The rms of the data from the orbital solution is 0.89 km\,s$^{-1}$
for Chiron, 1.39 for Asiago, 1.08 for \citet{1986AJ.....91..125K}, 1.25 for
\citet{2000AJ....119.1375F}, 1.79 for
\citet{1949ApJ...109...81S}, and 1.24 for \citet{1958ApJ...127..625K}.

We have performed rather extensive measurements of the emission lines
present on our Echelle spectra of T~CrB, recorded during SAP, hoping to be
able to trace or at least constraining the motion of the WD companion.  The
line profiles during SAP are however highly variable and complex, and so
different from line to line for the same orbital phase that no matter which
way and in how many components they are deconvolved, a clean and unambiguous
tracer of the WD motion could not be promptly identified.  Several and
different absorption and emission components superimpose along any given
line of sight, apparently more responsive to the disk and hot spot
brightness than to the orbital phase.  This is clearly demonstrated by the
atlases of line profiles during SAP, which are arranged according to the
orbital phase, that we present in Appendix~\ref{appx:atlasSAP} for the time
interval 2015-14-05 to 2023-06-01, and confirmed in
Appendix~\ref{appx:atlaspost} by the dense mapping in high resolution of the
H$\alpha$ profile that we have carried out at $\sim$1-week cadence since the
end of SAP, from 2023-06-01 to 2025-05-01.

The lines grow in complexity of the emission profile and in the multiplicity
of absorption components for orbital phases 0.4 to 0.8, i.e.  between the
passage of the RG at descending quadrature and its transit at inferior
conjunction.  The line-of-sight during this phase interval traverses the
densest and inner strata of the RG atmosphere, crosses the hot spot and the
inflated outer rim of the disk around it, aligns with the fraction of the
accretion stream that may continue past the hot spot, before finally
reaching the hottest inner radii of the disk: such a richness of emitting
and absorbing bodies aligned along the line-of-sight, each one animated by
largely different and continously varying radial velocities, clearly calls
for highly complex and lively lines profiles, and the atlases presented 
in Appendix~\ref{appx:atlasSAP} and Appendix~\ref{appx:atlaspost} are a
testimony to that.

The situation was apparently simpler during the long quiescence (not covered
by our observations that were triggered by T~CrB entering SAP), when
\citet{2004A&A...415..609S} were able to derive a good-looking sinusoidal
motion from the H$\alpha$ emission component in anti-phase with that of the
RG, clearly supporting a WD appreciably more massive than the RG.

 \begin{table*}[ht]
 \caption{Binary and disk parameters depending on WD mass and orbital
 inclination for the mass function $f(m)$=0.322 derived from orbital
 solution in Table~\ref{tab:orbit}.  Azimuths are angles in degrees counted
 from the line joining the two stars, positive in the direction of disk
 rotation.  M$_{\rm WD}$ and M$_{\rm RG}$ are the mass of the WD and RG, $i$
 and $a$ the orbital inclination and separation, R$_{\rm RG}$ is the radius
 of the Roche lobe.  $r_i$ and $V_i$ are the closest distance to WD and the
 velocity at the point of the accretion stream on a fly-by trajectory around
 the WD (thus $r_i$ is the minimum radius of the disk to intercept the
 stream, at $\sim$159$^\circ$ azimuth).  $r_c$, $\theta_c$, and $V_c$ are
 the disk's circularization radius and for the impacting stream the
 corresponding azimuth and velocity.  $r_o$, $\theta_o$, and $V_o$ are the
 corresponding values for the outer disk radius in the $r_o$=2$\times$$r_c$
 approximation.  Finally, the last column reports about eclipsing of the hot
 spot by the RG ('grz' standing for grazing condition).}
 \label{tab:disk}
 \centering
 \begin{tabular}{cccccccccccccccccc}
 \hline \hline
 &&&&&&\multicolumn{2}{c}{Intercept}&&\multicolumn{3}{c}{Circularization}&&\multicolumn{3}{c}{Outer}&HS\\ 
 \cline{7-8} \cline{10-12} \cline{14-16}  
M$_{\rm WD}$&M$_{\rm RG}$&$i$&$a$&R$_{\rm RG}$
&&$r_i$       & $V_i$ 
&& $r_c$      & $\theta_c$ & $V_c$ 
&&$r_o$       & $\theta_o$ & $V_o$ & eclip. \\ 
(M$_\odot$)&(M$_\odot$)&($^\circ$) &(R$_\odot$) &(R$_\odot$)
&&(R$_\odot$) &(km/s)
&&(R$_\odot$) &($^\circ$) &(km/s) 
&&(R$_\odot$) &($^\circ$) &(km/s) &\\
 \hline
        1.35 & 1.05 & 65.0 & 209.9 & 75.1 && 13.1 & 191.6 && 28.3 & 52.2 & 116.4 && 56.6 & 19.8 & 65.1 & yes   \\
        1.30 & 0.96 & 65.0 & 205.8 & 72.7 && 13.4 & 184.7 && 28.2 & 52.9 & 114.7 && 56.4 & 19.5 & 64.1 & yes   \\
        1.25 & 0.88 & 65.0 & 201.7 & 70.4 && 13.6 & 180.5 && 28.1 & 53.2 & 112.6 && 56.2 & 19.8 & 63.1 & yes   \\
 &&\\
        1.35 & 0.93 & 61.5 & 206.4 & 71.1 && 14.1 & 183.4 && 28.9 & 53.4 & 115.4 && 57.9 & 19.6 & 64.7 & grz   \\
        1.30 & 0.85 & 61.5 & 202.4 & 69.4 && 14.3 & 178.9 && 28.9 & 52.8 & 113.7 && 57.7 & 18.8 & 63.6 & grz   \\
        1.25 & 0.78 & 61.5 & 198.5 & 67.3 && 14.5 & 173.7 && 28.8 & 54.0 & 112.0 && 57.5 & 19.2 & 62.6 & grz   \\
 &&\\
	1.35 & 0.76 & 56.5 & 200.9 & 66.0 && 15.6 & 172.9 && 30.2 & 54.0 & 113.7 && 60.4 & 18.6 & 63.5 & no    \\
	1.30 & 0.69 & 56.5 & 197.1 & 63.8 && 15.7 & 169.5 && 30.2 & 52.2 & 111.1 && 60.4 & 17.6 & 62.2 & no    \\
	1.25 & 0.63 & 56.5 & 193.3 & 61.7 && 15.9 & 164.0 && 30.3 & 53.2 & 109.0 && 60.5 & 17.8 & 61.0 & no    \\
 \hline
 \end{tabular}
 \end{table*}

\section{Radiative modeling}
\label{sect:radiative}

The accurate $U$$B$$V$$R$$I$ photometry collected by ANS Collaboration and
covering both the quiescence and SAP phases allows to derive a detailed
radiative model of T~CrB.  In this paper we will limit our analysis of the
photometry to an assessment of the fraction of the Roche Lobe filled by the
RG, and to quantify the role and position of the hot spot in phase-reckoned
lightcurves.  The position relates to the azimuth, which in turn fixes the
outer radius of the disk; the latter governs the maximum orbital inclination
that still avoids eclipsing the hot spot, and ultimately affect the mass of
the red giant.

We adopt the same radiative modeling  carried out by
\citet{2023RNAAS...7..251M} in computing the synthetic lightcurve of the
secondary maximum during the 1946 outburst of T~CrB, and its description
will be not repeated here, except adding that for the radial dependence of
the temperature of the disk we followed the standard profile $T(r)=T_0
(R_{\rm WD}/r)^{3/4}(1- \sqrt{R_{\rm WD}/r})^{1/4}$
\citep[cf.][]{1995cvs..book.....W}, where $R_{\rm WD}$ and $M_{\rm WD}$ are
the radius and mass of the WD and $T_0=[(3GM_{\rm WD}\dot{M})/(8\pi\sigma
R^{3}_{\rm WD})]^{1/4}$ relates to the $\dot{M}$ mass accretion flow through
the disk.  The masses of the RG and WD, and the orbital inclination, have to
satisfy the mass function 0.322 derived in the orbital solution of
Table~\ref{tab:orbit}, with the orbital period and zero phase taken verbatim
from it.

   \begin{figure}[ht!]
   \centering
   \includegraphics[width=\hsize]{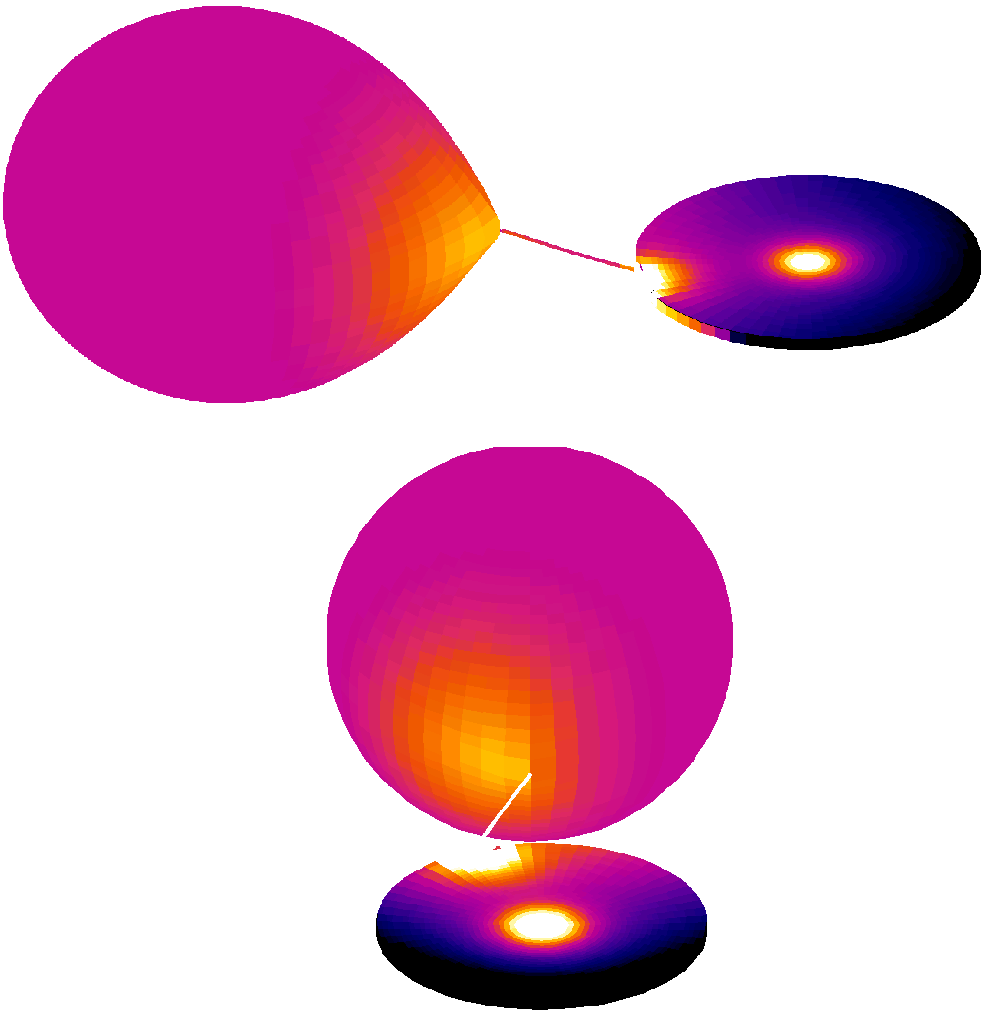}
      \caption{3D-views of T~CrB in quiescence from the radiative 
      modeling carried out in sect.~\ref{sect:radiative}, for the 
      RG filling completely its Roche lobe, the disk extending to the
      standard outer radius of twice its circularization value, an
      orbital inclination of 61$^{\circ}_{.}5$, and masses of the WD
      and the RG of 1.35 and 0.93 M$_\odot$, respectively (cf.
      Tab.~\ref{tab:disk}). The stretching of the black-to-white 
      color palette is different for the RG and for the disk + hot 
      spot, respectively 3450$-$3600~K and 1400$-$1900~K.}
         \label{fig:view3D}
   \end{figure}

   \begin{figure}[ht!]
   \centering
   \includegraphics[width=\hsize]{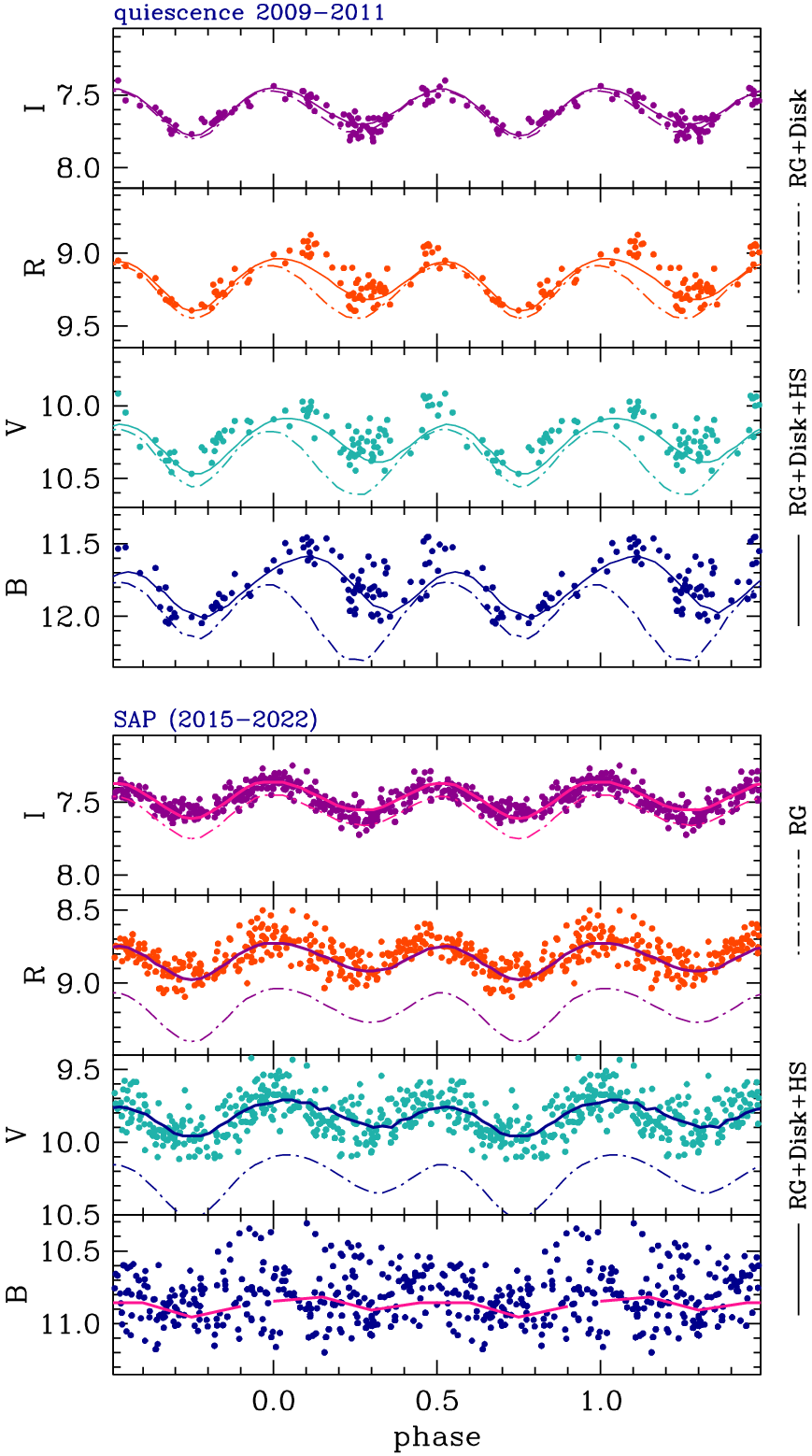}
      \caption{The lightcurve of T CrB, separately for quiescence and
      SAP, is fitted with the radiative modeling described in 
      sect.~\ref{sect:radiative} and the same parameters leading up to 
      the 3D view of Fig.~\ref{fig:view3D}. In the panels for quiescence
      at the top, the two sets of fitting curves differ by the inclusion
      of the hot spot (following the parameters listed in
      sect.~\ref{sect:radiative}); in the panels for SAP at the bottom,
      the solid curve is the fitting considering the full system, while the 
      dot-dashed curve isolates the contribution of the sole RG.}
         \label{fig:model}
   \end{figure}

A sample representation of the 3D view produced by our radiative modeling is
shown in Fig.~\ref{fig:view3D}, relative to T~CrB during the quiescence
phase prior to SAP, and separately for the passage of the RG at superior
conjunction (phase 0.25) and descending quadrature (phase 0.5), for the
parameters on the 4th line from top of Table~\ref{tab:disk}, those providing
the best fit to photometric lightcurves.  The computed synthetic lightcurves
are compared to the observed ones in Fig.~\ref{fig:model}, separately for
quiescence (top panels) and SAP phases (bottom panels).  Concerning the main
goals stated the opening of this section, the results in
Fig.~\ref{fig:model} indicates that:

\begin{itemize}

	\item the RG fills completely its Roche lobe; even a modest under-filling 
by 3\% would cause a drop by 0.10~mag of the overall brightness, incompatible
with Fig.~\ref{fig:model}. \citet{2023AstL...49..501M} concluded the same
from modeling in the infrared;

	\item the quiescent lightcurve constrains the azimuth of the hot
spot (HS) to be 20$\pm$10$^\circ$; the effective angular dimension of the HS
as seen from the WD is about 23$^\circ$, with a peak temperature of 5910~K
declining to 5080~K at the HS periphery;

	\item the low temperature of the HS precludes it from being an
important site of formation for HeII emission lines, in agreement with the
results from the Doppler tomography performed by \citet{2025A&A...694A..85P}
that favors a formation in the inner parts of the disk;

	\item coupled with the dynamic trajectory of the accretion stream,
the azimuth of HS implies that the outer radius $r_o$ of the disk extends to
twice the circularization radius $r_c$, as expected in a standard disk
\citep{1995cvs..book.....W}.  An outer radius equal to $r_c$ would move the
azimuth of the HS to $\sim$53$^\circ$, and to $\sim$159$^\circ$ if equal the
minimum represented by the intercept radius $r_i$ (cf. 
Table~\ref{tab:disk});

	\item the HS irradiates the facing side of the RG, rising the surface
temperature from 3500~K to a peak value of 3560~K reached at the sub-HS
point (yellowish color in Fig.~\ref{fig:view3D}).  It also irradiates the
closer portion of the disk, slightly rising the local temperature above that
corresponding to the pure accretion flow (purple-to-pink color in
Fig.~\ref{fig:view3D});

	\item no matter the actual value of $\dot{M}$, in quiescence the
outer parts of the disk remain much cooler than the surface of the RG and
completely neutral: in spite of their large angular extent (similar to that
of the RG), they do not appreciably contribute to the overall brightness of
T~CrB at optical wavelengths;

	\item  the HS contributes significantly to the overall brightness of
T~CrB in quiescence even in $V$ and $R$ bands.  This fact strongly supports
the counter-intuitive results by \citet{2021MNRAS.505.6121M} and
\citet{2024A&A...683A..84M} for the detectability of flickering in symbiotic
stars (SySts).  \citet{2024A&A...683A..84M} was able to isolate flickering
signatures from {\it TESS} ultra-precise observations of SySts from space,
in spite of the fact that {\it TESS} observes in white-light over the entire
optical range and therefore should be vastly dominated by the direct
emission from the RG.  Thanks to highly-accurate, ground-based photometry in
the $B$ band, \citet{2021MNRAS.505.6121M} have been able to detect
flickering at a few milli-mag level from SySts accreting at such low levels
that their nature is betrayed only by a faint emission in H$\alpha$ on
high-resolution spectra, which becomes visible only {\it after} the
subtraction of a template spectrum for the RG;

	\item the HS plays such an important role in shaping the lightcurve
of T~CrB that any change in its brightness or in the radius of the disk (and
therefore in the azimuth of HS) would significantly alter the times of
passage at minimum system brightness; HS variability developing on long time
scales, as those reported by \citet{2004A&A...415..609S}, could 
impact attempts to detect a change in the orbital period by comparing
minima in the lightcurves before and after the 1946 outburst;

	\item a large amplitude flickering has been observed at each of the
many visits paid to T~CrB during quiescence, implying that the HS is not
eclipsed.  The maximum orbital inclination that still avoids eclipses of the
HS by the RG is 61.5$^\circ$.  From the 0.322 mass function and for a WD
mass of $\sim$1.35~M$_\odot$ (as generally adopted in theoretical
models of T~CrB), such an orbital inclination returns a mass of
0.93~M$_\odot$ for the RG (cf.  Table~\ref{tab:disk}), well below the
typical mass for field, single M3III giant
\citep[eg.][]{2000asqu.book.....C}.  Given the Roche-lobe filling status of
the RG, much of the missing mass could have been transferred to the WD, and
there retained to favor its growth toward the Chandrasekhar limit.  In
accordance with the high retention efficiency ([$M_{\rm acc} - M_{\rm
ej}]/M_{\rm acc}$) recently discovered by \citet{2025ApJ...982...89S}, the
many nova outbursts that T~CrB has underwent in its life have evidently not
altered the mass-gaining path followed by the WD;

	\item the accretion disk, which remains faint during quiescence,
dominates instead the lightcurve during SAP, contributing significantly even
in the $I$ band to the overall brightness; during SAP the relative
importance of the HS on the system brightness is greatly diminished compared
to quiescence.  An HS similar to quiescence may still be present during SAP,
as expected considering that the RG keeps spilling mass via $L_1$ even
during SAP.  The fact however that during SAP the HS does {\it not} turn
brighter than quiescence, implies that SAP is not triggered by an increased
mass-loss rate of the RG but is instead caused by some radial collapse of
the disk that enhances the mass-flow toward the innermost radii and
ultimately toward the WD;

	\item in this interpretation, the initial $\sim$70 yrs of the
$\sim$80 yrs intra-outburst period are primarily spent in building up mass
in the disk, mass that is then flushed toward the WD primarily during the
last $\sim$10 yrs of the intra-outburst period, giving rise to SAP-like events;

	\item the radial collapse of the disk seems to propagate to the
outer regions at later epochs.  In fact, the $I$-band lightcurve on the
right panel of Fig.~\ref{fig:UBVRI} maintains the SAP level well through the
deep-minimum phase, until mid 2024 when it drops to the value of the bare
RG, simultaneously with the reprisal in brightness at shorter wavelengths
powered by a regain in the mass flow at inner radii.

\end{itemize}

\section{Rotational velocity}\label{sec:Rotation}

\citet{2006MNRAS.365.1215Z, 2007MNRAS.380.1053Z} derived the rotational
velocity of the RG in symbiotic stars from the width of absorption lines
measured on {\sc FEROS} spectra obtained at 48,000 resolving power.  They
inferred that $\sim$90\% of the S-type RGs have $V_{rot} \sin i$ in the
interval 4.5 $-$ 11.7~km\,s$^{-1}$, rotate faster than single field giants
of similar spectral type, and are corotating (synchronized) with the orbital
revolution.  From their infrared spectra recorded at 44,000 resolving power,
\citet{2004IAUS..215..168F} derived for the RG of T~CrB a rotational
velocity $V_{rot} \sin i$=5.1~km\,s$^{-1}$, while
\citet{2025arXiv250220664H} obtained $V_{rot} \sin i$=8.7~km\,s$^{-1}$ from
optical spectra at a low 15,000 resolving power.

Our Chiron spectrum for 2022-08-16 has been recorded at a much higher
resolving power than either \citet{2004IAUS..215..168F} and
\citet{2025arXiv250220664H}, allowing us to derive a more accurate value of
$V_{rot} \sin i$.  For an instrumental line-spread function (LSF) with a
Gaussian shape and for low rotational velocity, the observed width of
stellar lines can be expressed as:
\begin{equation}
\sigma^{2}_{\rm obs} = (\sigma_{\rm rot}\sin i)^{2} + \sigma^{2}_{\rm LSF} +
\sigma^{2}_{\xi}
\end{equation}
where $\sigma_{\xi}$ is the microturbolent velocity.  For the M3III giant in
T~CrB \citet{2008PASP..120..492W} derived $\sigma_{\xi}$=2~km\,s$^{-1}$.
 
Isolated telluric lines provide an efficient way to estimate $\sigma_{\rm
LSF}$ because they illuminate the spectrograph slit in exactly the same way
as the stellar seeing disk, providing a more accurate estimate than using
lines from the comparison lamp or the sky background, which illuminates the
slit uniformly.  Usually $\sigma_{\rm LSF}$ degrades moving away from the
center of the spectrograph's optical path because of aberrations, so to
exploit the best Chiron resolving power we looked for the telluric O$_2$
band centered at 6278~\AA\ that is recorded by Chiron close to the center of
central order N.48.  From seven unbleded lines belonging to the O$_2$
telluric band we derived $\sigma_{\rm LSF}$=0.0306$\pm$0.0030 corresponding
to a local resolving power of $\lambda/\Delta \lambda$=87,200.

The width of stellar lines $\sigma_{\rm obs}$ must obviously be measured at the
same position of the telluric lines used to estimate $\sigma_{\rm LSF}$ to
exploit similar optical performance by the spectrograph.  From ten isolated
stellar lines interspersed with the O$_2$ telluric ones we derive
$\sigma_{\rm obs}$=0.1124$\pm$0.0089, from which we finally obtain: $V_{\rm
rot} \sin i$=4.75$\pm$0.26 km\, s$^{-1}$.  For an inclination of
$i$=61.5$^\circ$, such velocity corresponds to corotation at a stellar
radius of 24.3~R$_\odot$.  From the radiative modeling in
sect.~\ref{sect:radiative}, the radius of the RG filling its Roche lobe is
$\sim$71~R$_\odot$ (cf.  the fifth column of Table~\ref{tab:disk}), for
which the corotation velocity would be V$_{\rm rot}$=16.0~km\,s$^{-1}$.  The
RG in T CrB is therefore far from corotation conditions, as if the outer
layers of the RG are removed via spilling through $L_1$ at a rate faster
than it is required to achieve corotation. In fact, the theoretical 
time-scale for synchronization would be just a few hundreds years following 
\citet{2007MNRAS.380.1053Z}, orders of magnitude shorter that the time spent
by T~CrB is the symbiotic mass-transfer phase.

\section{Conclusions}

The large body of multi-wavelength observations assembled for T~CrB both
during the quiescence following the 1946 outburst and in particular during
the recent 2015-2023 enhanced mass-transfer phase, together with
the results of orbital solution and radiative modeling, have allowed to draw
a consistent picture of T~CrB and how its WD accretes the mass that is
required to trigger the outburst.

The RG completely fills its Roche lobe and it is not corotating,  a possible
consequence of the continuous removal of its outer layers by the incessant
spilling through $L_1$.  The material lost by the RG goes to form a disk
around the WD.  The $\sim$20$^\circ$ azimuth of the bright hot spot formed
by the impacting accretion stream fixes the outer radius of the disk to
$\sim$58~R$_\odot$, twice the circularization radius as in the standard
theory of accretion disks.  The hot spot is not eclipsed, and its 23$^\circ$
angular extension and mild temperature (5900~K degrading to 5100~K at the
periphery) make it well visible in the lightcurve of T~CrB even at $V$ and
$R$ bands.  For a WD of 1.35~M$_\odot$, the orbital and radiative modeling
return an inclination of 61$^\circ$ and a mass of 0.93~M$_\odot$ for the RG,
making the mass-transfer dynamically stable. The RG has been
stripped out of the outer 0.3-0.5 solar masses, exposing to mass-transfer
toward the WD the internal layers which composition (chemical and isotopic)
should significantly differ from the Solar mixture generally assumed 
in theoretical models of the outburst.

The hot spot did not brighten during the enhanced mass-transfer phase,
indicating that SAP is not triggered by an increase in the mass loss from
the RG but is instead caused by the inside-out collapse of the accretion
disk grown in mass beyond the stable configuration.  The disk seems spending
the initial $\sim$70 of the $\sim$80 yrs intra-outburst period mainly to
grown in mass, transferring to the WD only a limited amount of what it is
gaining; it is during the last $\sim$10 of the $\sim$80 yrs intra-outburst
interval that most of the mass-transfer to the WD takes place, in a 4:1
proportion and at mean rate $\sim$28 faster than in quiescence.

Comparing with the enhanced mass-transfer phase that preceded the 1946
outburst, the present one attained a lower brightness, suggesting that a
lower amount of mass has been transferred from the disk to the WD.  This
could justify the delay affecting the outburst, not yet occurred two full
years past the end of the enhanced mass-transfer phase while in 1946 it
followed just 6 months after.

While the inner parts of the disk emptied during the enhanced mass-transfer
phase which ended in late April 2023, the outer parts started migrating
inward, and from $\sim$May 2024 begun restoring the mass-flow at shorter
radii with the noticeable effect that the brightness of T~CrB in the optical
and ultraviolet rose back to levels higher than in quiescence.  The
consequent revamp in the accretion rate toward the WD could fill any deficit
left over by the under-luminous 2015-2023 enhanced mass-transfer phase, and
ultimately lead the shell accumulated on the surface of the WD to reach
ignition conditions.  The material restoring the mass-flow at the inner
radii of the disk migrated inward from the outer parts of the disk which
radiates primarily in the far red/IR: in fact, simultaneous with the
re-brightening in the blue and UV that started in May 2024, T~CrB became
fainter in the $I$ band almost down to the brightness of the bare red giant.

\section{Data availability}

Tables 1 and A.1 are only available in electronic form at the CDS via
anonymous ftp to cdsarc.u-strasbg.fr (130.79.128.5) or via
http://cdsweb.u-strasbg.fr/cgi-bin/qcat?J/A+A/.  The spectroscopic data
presented in this paper are also available via CDS.

\begin{acknowledgements}
We acknowledge the anonymous Referee for useful and constructive comments.
This work has been in part supported by INAF 2023 MiniGrant Program
(contract C93C23008470001 to UM).  NM acknowledges financial support through
ASI-INAF and ’Mainstream’ agreement 2017-14-H.0 (PI: T.  Belloni).  This
work has made use of data provided by Digital Access to a Sky Century @
Harvard (DASCH), which has been partially supported by NSF grants
AST-0407380, AST-0909073, and AST-1313370.  Work on DASCH Data Release 7
received support from the Smithsonian American Women’s History Initiative
Pool.  Part of this work is based on archival data, software or online
services provided by the Space Science Data Center - ASI.
\end{acknowledgements}

\bibliographystyle{aa}
\bibliography{paper_rev.bib}

\begin{appendix}

\onecolumn
\section{Swift UVOT observations of T~CrB}

In order to trace the evolution of T~CrB in the near-UV, we selected a series of pointings
performed over nearly two decades, between 2008 and 2025, with the UltraViolet Optical Telescope (UVOT; Roming et al.
2005) onboard the Neil Gehrels {\it Swift} satellite (Gehrels et al. 2004), which
observed in the {\it UVW1} (2600 \AA), {\it UVM2} (2246 \AA), and {\it UVW2}
(1928 \AA) filters. See sect.~\ref{sect:Swift} for details about the
measurements of these data.

\begin{longtable}{cccrccccrrr}
\caption{Archive Swift UVOT observations of T~CrB.}\\
\label{tab:UVOT} 
  Obs. ID   &   Date       &  UT start   &   expt     &        W2         &       M2         &         W1       &&          W2      &        M2         &          W1       \\ \cline{5-7} \cline{9-11}
            &              &             &   (sec)    & \multicolumn{3}{c}{(Vega system magnitudes)}            &&   \multicolumn{3}{c}{(fluxes $\times$10$^{-15}$cm$^{-2}$s$^{-1}$\AA$^{-1}$)}\\ 
\hline
\endfirsthead
\caption{continued.}\\
  Obs. ID   &   Date       &  UT start   &   expt     &        W2         &       M2         &         W1       &&          W2      &        M2         &          W1       \\ \cline{5-7} \cline{9-11}
            &              &             &   (sec)    & \multicolumn{3}{c}{(Vega system magnitudes)}            &&   \multicolumn{3}{c}{(fluxes $\times$10$^{-15}$cm$^{-2}$s$^{-1}$\AA$^{-1}$)}\\ 
\hline
\endhead
\hline
00035171006 &  2008-03-16  &  20:32:24   &    269     &  14.44$\pm$0.04   &                  &                  &&    9.0$\pm$0.3   &                   &                   \\
00035171006 &  2008-03-16  &  20:41:41   &    255     &                   &                  & 13.67$\pm$0.04   &&                  &                   &   13.6$\pm$0.5    \\
00035171006 &  2008-03-16  &  20:37:03   &    269     &                   & 14.64$\pm$0.04   &                  &&                  &    6.5$\pm$0.2    &                   \\
00035171006 &  2008-03-17  &  02:56:24   &    269     &  14.38$\pm$0.04   &                  &                  &&    9.5$\pm$0.3   &                   &                   \\
00035171006 &  2008-03-17  &  03:01:02   &    269     &                   & 14.82$\pm$0.05   &                  &&                  &   5.47$\pm$0.18   &                   \\
00035171006 &  2008-03-17  &  03:03:41   &    255     &                   &                  & 13.87$\pm$0.04   &&                  &                   &   11.3$\pm$0.4    \\
00045776001 &  2011-11-27  &  12:09:33   &    542     &  14.17$\pm$0.04   &                  &                  &&   11.5$\pm$0.4   &                   &                   \\
00045776001 &  2011-11-27  &  12:18:49   &    542     &                   & 14.84$\pm$0.04   &                  &&                  &   5.36$\pm$0.15   &                   \\
00045776001 &  2011-11-27  &  12:28:06   &    465     &                   &                  & 13.69$\pm$0.04   &&                  &                   &   13.4$\pm$0.5    \\
00045776001 &  2011-11-28  &  02:50:12   &    364     &                   & 14.96$\pm$0.04   &                  &&                  &   4.82$\pm$0.15   &                   \\
00045776001 &  2011-11-28  &  02:43:57   &    364     &  14.55$\pm$0.04   &                  &                  &&    8.1$\pm$0.3   &                   &                   \\
00045776001 &  2011-11-28  &  02:56:27   &    385     &                   &                  & 13.79$\pm$0.04   &&                  &                   &   12.1$\pm$0.5    \\
00045776002 &  2011-12-04  &  09:38:49   &    363     &                   &                  & 13.56$\pm$0.04   &&                  &                   &   15.1$\pm$0.6    \\
00045776002 &  2011-12-04  &  09:24:23   &    421     &  14.10$\pm$0.04   &                  &                  &&   12.3$\pm$0.4   &                   &                   \\
00045776002 &  2011-12-04  &  09:31:36   &    421     &                   & 14.60$\pm$0.04   &                  &&                  &   6.72$\pm$0.19   &                   \\
00081659002 &  2015-09-24  &  02:37:18   &   1632     &  10.13$\pm$0.10   &                  &                  &&    470$\pm$50    &                   &                   \\
00045776004 &  2015-10-01  &  01:01:23   &    151     &   9.92$\pm$0.12   &                  &                  &&    570$\pm$70    &                   &                   \\
00045776005 &  2017-01-18  &  04:15:03   &   1174     &                   &  9.26$\pm$0.10   &                  &&                  &    920$\pm$90     &                   \\
00045776005 &  2017-01-18  &  05:49:17   &   1219     &                   &  9.30$\pm$0.10   &                  &&                  &    880$\pm$90     &                   \\
00045776005 &  2017-01-18  &  07:25:17   &   1219     &                   &  9.28$\pm$0.10   &                  &&                  &    900$\pm$90     &                   \\
00045776005 &  2017-01-18  &  10:41:22   &    919     &                   &  9.34$\pm$0.10   &                  &&                  &    860$\pm$90     &                   \\
00045776005 &  2017-01-18  &  11:58:48   &   1366     &                   &  9.30$\pm$0.10   &                  &&                  &    880$\pm$90     &                   \\
00045776005 &  2017-01-18  &  13:34:15   &   1457     &                   &  9.24$\pm$0.10   &                  &&                  &    930$\pm$90     &                   \\
00045776005 &  2017-01-18  &  15:17:25   &    916     &                   &  9.30$\pm$0.10   &                  &&                  &    880$\pm$90     &                   \\
00045776005 &  2017-01-18  &  17:00:24   &    975     &                   &  9.35$\pm$0.10   &                  &&                  &    850$\pm$90     &                   \\
00045776005 &  2017-01-18  &  20:21:06   &    580     &                   &  9.40$\pm$0.11   &                  &&                  &    810$\pm$90     &                   \\
00045776007 &  2017-02-08  &  13:41:45   &    247     &                   &  9.39$\pm$0.11   &                  &&                  &    820$\pm$90     &                   \\
00045776007 &  2017-02-08  &  21:58:15   &    217     &                   &  9.41$\pm$0.11   &                  &&                  &    800$\pm$90     &                   \\
00045776026 &  2017-09-26  &  02:03:43   &    662     &                   &  9.63$\pm$0.11   &                  &&                  &    650$\pm$70     &                   \\
00045776026 &  2017-09-26  &  10:01:44   &   1605     &                   & 9.53$\pm$0.10    &                  &&                  &    710$\pm$70     &                   \\
00045776026 &  2017-09-26  &  16:41:57   &    708     &                   & 9.55$\pm$0.10    &                  &&                  &    700$\pm$70     &                   \\
00045776026 &  2017-09-26  &  21:24:02   &    585     &                   & 9.57$\pm$0.11    &                  &&                  &    690$\pm$80     &                   \\
00045776026 &  2017-09-26  &  23:15:14   &    395     &                   & 9.48$\pm$0.11    &                  &&                  &    740$\pm$80     &                   \\
00045776027 &  2017-10-27  &  10:44:33   &    730     &                   & 9.26$\pm$0.10    &                  &&                  &    920$\pm$90     &                   \\
00045776032 &  2018-07-27  &  03:37:45   &    541     &                   & 10.60$\pm$0.04   &                  &&                  &    266$\pm$6      &                   \\
00011548001 &  2019-09-04  &  02:34:46   &    127     &                   & 10.18$\pm$0.13   &                  &&                  &    400$\pm$50     &                   \\
00011548001 &  2019-09-04  &  08:55:36   &    138     &                   &  9.62$\pm$0.12   &                  &&                  &    660$\pm$80     &                   \\
00011548001 &  2019-09-04  &  15:28:21   &     93     &                   & 10.05$\pm$0.13   &                  &&                  &    440$\pm$60     &                   \\
00013922001 &  2020-12-06  &  00:51:42   &    191     &                   & 10.71$\pm$0.04   &                  &&                  &    241$\pm$6      &                   \\
00012011005 &  2021-07-10  &  08:47:26   &    147     &                   & 10.89$\pm$0.04   &                  &&                  &    205$\pm$5      &                   \\                                 
00012011006 &  2021-07-10  &  09:02:03   &     51     &  10.99$\pm$0.04   &                  &                  &&    214$\pm$8     &                   &                   \\
00012011013 &  2021-08-24  &  02:28:32   &    200     &                   & 10.76$\pm$0.04   &                  &&                  &    231$\pm$5      &                   \\
00013558010 &  2022-03-08  &  14:05:24   &    149     &                   & 10.60$\pm$0.04   &                  &&                  &    266$\pm$6      &                   \\
00013558017 &  2022-06-22  &  07:49:39   &    193     &                   & 10.67$\pm$0.04   &                  &&                  &    249$\pm$6      &                   \\
00013558035 &  2022-12-10  &  19:22:45   &    124     &                   &                  & 11.14$\pm$0.04   &&                  &                   &  140$\pm$5        \\
00013558035 &  2022-12-10  &  19:27:11   &    249     &  11.09$\pm$0.04   &                  &                  &&    196$\pm$6     &                   &                   \\
00013558035 &  2022-12-10  &  19:32:35   &    197     &                   & 11.23$\pm$0.04   &                  &&                  &    149$\pm$3      &                   \\
00013558041 &  2023-01-10  &  17:56:20   &    153     &                   & 10.74$\pm$0.04   &                  &&                  &    235$\pm$6      &                   \\
00013558053 &  2023-03-10  &  16:20:13   &    160     &                   & 10.66$\pm$0.04   &                  &&                  &    253$\pm$6      &                   \\
00013558067 &  2023-05-12  &  19:00:09   &    154     &  11.20$\pm$0.04   &                  &                  &&    178$\pm$6     &                   &                   \\
00013558067 &  2023-05-12  &  19:03:33   &     82     &                   & 11.27$\pm$0.04   &                  &&                  &    144$\pm$4      &                   \\
00013558072 &  2023-05-25  &  23:11:29   &    126     &                   &                  & 11.31$\pm$0.04   &&                  &                   &   119$\pm$5       \\
00013558072 &  2023-05-25  &  23:15:59   &    252     &  11.30$\pm$0.04   &                  &                  &&    161$\pm$4     &                   &                   \\
00013558072 &  2023-05-25  &  23:21:28   &    145     &                   & 11.49$\pm$0.04   &                  &&                  &    118$\pm$3      &                   \\
00013558078 &  2023-06-09  &  10:37:24   &    130     &                   &                  & 11.78$\pm$0.04   &&                  &                   &    78$\pm$3       \\
00013558078 &  2023-06-09  &  10:42:02   &    261     &  11.64$\pm$0.04   &                  &                  &&    118$\pm$4     &                   &                   \\
00013558078 &  2023-06-09  &  10:47:41   &    191     &                   & 11.85$\pm$0.04   &                  &&                  &     85$\pm$2      &                   \\
00013558093 &  2023-07-09  &  17:21:09   &    223     &                   & 12.66$\pm$0.04   &                  &&                  &   40.0$\pm$1.0    &                   \\
00013558092 &  2023-07-09  &  23:29:46   &    127     &                   &                  & 12.17$\pm$0.04   &&                  &                   &    54$\pm$2       \\
00013558092 &  2023-07-09  &  23:34:17   &    215     &  12.37$\pm$0.04   &                  &                  &&   60.1$\pm$1.8   &                   &                   \\
00013558107 &  2023-08-08  &  10:56:08   &    119     &                   &                  & 12.88$\pm$0.04   &&                  &                   &  28.0$\pm$1.1     \\
00013558107 &  2023-08-08  &  11:00:23   &    238     &  13.45$\pm$0.04   &                  &                  &&   22.4$\pm$0.7   &                   &                   \\
00013558107 &  2023-08-08  &  11:05:33   &    140     &                   & 13.47$\pm$0.04   &                  &&                  &   18.9$\pm$0.6    &                   \\
00013558121 &  2023-09-05  &  01:50:33   &     79     &                   &                  & 12.97$\pm$0.04   &&                  &                   &  25.8$\pm$1.1     \\
00013558121 &  2023-09-05  &  01:53:27   &    157     &  13.48$\pm$0.04   &                  &                  &&   21.6$\pm$0.8   &                   &                   \\
00013558121 &  2023-09-05  &  01:56:56   &    117     &                   & 13.78$\pm$0.05   &                  &&                  &   14.3$\pm$0.5    &                   \\
00013558139 &  2023-10-03  &  18:01:39   &     85     &                   &                  & 13.13$\pm$0.04   &&                  &                   &  22.2$\pm$0.9     \\
00013558139 &  2023-10-03  &  18:04:46   &    171     &  13.39$\pm$0.04   &                  &                  &&   23.6$\pm$0.8   &                   &                   \\
00013558139 &  2023-10-03  &  18:08:32   &    142     &                   & 13.43$\pm$0.04   &                  &&                  &   19.8$\pm$0.5    &                   \\
00013558155 &  2023-10-28  &  18:17:34   &    139     &                   &                  & 13.22$\pm$0.04   &&                  &                   &  20.5$\pm$0.8     \\
00013558155 &  2023-10-28  &  18:20:07   &    137     &  13.19$\pm$0.04   &                  &                  &&   28.4$\pm$0.8   &                   &                   \\
00013558155 &  2023-10-28  &  18:23:10   &    104     &                   & 13.48$\pm$0.05   &                  &&                  &   18.8$\pm$0.6    &                   \\
00013558161 &  2023-11-28  &  14:27:15   &    130     &                   &                  & 13.04$\pm$0.04   &&                  &                   &  24.2$\pm$0.9     \\
00013558161 &  2023-11-28  &  14:31:52   &    260     &  13.39$\pm$0.04   &                  &                  &&   23.7$\pm$0.6   &                   &                   \\
00013558161 &  2023-11-28  &  14:37:32   &    200     &                   & 13.56$\pm$0.04   &                  &&                  &   17.4$\pm$0.5    &                   \\
00013558167 &  2023-12-12  &  21:58:16   &    133     &                   &                  & 12.82$\pm$0.04   &&                  &                   &  29.7$\pm$1.1     \\
00013558167 &  2023-12-12  &  22:02:59   &    265     &  13.31$\pm$0.04   &                  &                  &&   25.3$\pm$0.8   &                   &                   \\
00013558167 &  2023-12-12  &  22:08:43   &    189     &                   & 12.71$\pm$0.04   &                  &&                  &   38.1$\pm$0.9    &                   \\
00013558173 &  2023-12-26  &  19:40:53   &    104     &                   &                  & 12.44$\pm$0.04   &&                  &                   &  42.3$\pm$1.6     \\
00013558173 &  2023-12-26  &  19:44:40   &    208     &  13.02$\pm$0.04   &                  &                  &&   33.1$\pm$1.1   &                   &                   \\
00013558173 &  2023-12-26  &  19:49:14   &    159     &                   & 13.09$\pm$0.04   &                  &&                  &   27.0$\pm$0.7    &                   \\
00013558185 &  2024-01-23  &  01:31:24   &    129     &                   &                  & 11.96$\pm$0.04   &&                  &                   &    66$\pm$3       \\
00013558185 &  2024-01-23  &  01:35:58   &    257     &  12.48$\pm$0.04   &                  &                  &&   54.4$\pm$1.7   &                   &                   \\
00013558185 &  2024-01-23  &  01:41:33   &    199     &                   & 12.59$\pm$0.04   &                  &&                  &   42.6$\pm$1.0    &                   \\
00013558193 &  2024-02-06  &  18:32:37   &     78     &                   &                  & 12.72$\pm$0.04   &&                  &                   &  32.5$\pm$1.3     \\
00013558193 &  2024-02-06  &  18:38:52   &    121     &                   & 13.01$\pm$0.04   &                  &&                  &   29.1$\pm$0.8    &                   \\
00013558193 &  2024-02-06  &  18:35:27   &    155     &  12.81$\pm$0.04   &                  &                  &&   40.3$\pm$1.3   &                   &                   \\
00097564005 &  2024-04-11  &  17:14:59   &    173     &                   &                  & 13.33$\pm$0.04   &&                  &                   &  18.5$\pm$0.7     \\
00097564006 &  2024-04-11  &  17:33:32   &     82     &  13.87$\pm$0.05   &                  &                  &&   15.2$\pm$0.6   &                   &                   \\
00097564004 &  2024-04-11  &  20:24:59   &    882     &                   & 13.79$\pm$0.04   &                  &&                  &   14.2$\pm$0.3    &                   \\
00097564020 &  2024-05-11  &  11:37:11   &    177     &                   &                  & 11.59$\pm$0.04   &&                  &                   &    92$\pm$3       \\
00097564020 &  2024-05-11  &  11:32:36   &    266     &                   & 11.84$\pm$0.04   &                  &&                  &     85$\pm$2      &                   \\
00097564020 &  2024-05-11  &  11:41:53   &    298     &  11.93$\pm$0.04   &                  &                  &&     90$\pm$3     &                   &                   \\
00097564039 &  2024-06-10  &  19:10:07   &    276     &                   & 13.34$\pm$0.04   &                  &&                  &   21.3$\pm$0.5    &                   \\
00097564039 &  2024-06-10  &  19:14:53   &    184     &                   &                  & 12.89$\pm$0.04   &&                  &                   &  27.8$\pm$1.1     \\
00097564039 &  2024-06-10  &  19:19:45   &    306     &  13.37$\pm$0.04   &                  &                  &&   24.1$\pm$0.7   &                   &                   \\
00097564057 &  2024-07-10  &  10:58:33   &    258     &                   & 11.95$\pm$0.04   &                  &&                  &   77.0$\pm$1.9    &                   \\
00097564057 &  2024-07-10  &  11:03:00   &    171     &                   &                  & 11.73$\pm$0.04   &&                  &                   &    81$\pm$3       \\
00097564057 &  2024-07-10  &  11:07:32   &    318     &  11.64$\pm$0.04   &                  &                  &&    118$\pm$4     &                   &                   \\
00097564075 &  2024-08-09  &  19:27:38   &    273     &                   & 11.46$\pm$0.04   &                  &&                  &    121$\pm$3      &                   \\
00097564075 &  2024-08-09  &  19:32:21   &    182     &                   &                  & 11.17$\pm$0.04   &&                  &                   &   136$\pm$5       \\
00097564075 &  2024-08-09  &  19:37:09   &    341     &  11.52$\pm$0.04   &                  &                  &&    132$\pm$4     &                   &                   \\
00097564094 &  2024-09-08  &  06:51:45   &     69     &  11.19$\pm$0.04   &                  &                  &&    179$\pm$6     &                   &                   \\
00097564096 &  2024-09-08  &  10:26:44   &    276     &                   & 11.03$\pm$0.04   &                  &&                  &    179$\pm$4      &                   \\
00097564120 &  2024-10-08  &  01:23:36   &    329     &                   & 11.48$\pm$0.04   &                  &&                  &    119$\pm$3      &                   \\
00097564120 &  2024-10-08  &  01:29:16   &    219     &                   &                  & 11.30$\pm$0.04   &&                  &                   &   120$\pm$5       \\
00097564120 &  2024-10-08  &  01:35:02   &    407     &  11.35$\pm$0.04   &                  &                  &&    154$\pm$4     &                   &                   \\
00097564137 &  2024-11-29  &  23:17:03   &    297     &                   & 12.15$\pm$0.04   &                  &&                  &   64.0$\pm$1.6    &                   \\
00097564137 &  2024-11-29  &  23:22:10   &    198     &                   &                  & 11.97$\pm$0.04   &&                  &                   &    65$\pm$2       \\
00097564137 &  2024-11-29  &  23:27:23   &    327     &  12.06$\pm$0.04   &                  &                  &&     80$\pm$2     &                   &                   \\
00097564143 &  2024-12-09  &  07:32:48   &    259     &                   & 12.00$\pm$0.04   &                  &&                  &   73.6$\pm$1.8    &                   \\
00097564143 &  2024-12-09  &  07:37:16   &    172     &                   &                  & 11.80$\pm$0.04   &&                  &                   &    76$\pm$3       \\
00097564143 &  2024-12-09  &  07:41:50   &    183     &  12.10$\pm$0.04   &                  &                  &&     78$\pm$2     &                   &                   \\
00097564161 &  2025-01-08  &  04:17:02   &    315     &                   & 11.30$\pm$0.04   &                  &&                  &    140$\pm$3      &                   \\
00097564161 &  2025-01-08  &  04:22:27   &    209     &                   &                  & 11.09$\pm$0.04   &&                  &                   &   146$\pm$6       \\
00097564161 &  2025-01-08  &  04:27:58   &    411     &  11.22$\pm$0.04   &                  &                  &&    174$\pm$6     &                   &                   \\
00097564168 &  2025-01-18  &  08:54:10   &    259     &                   & 11.22$\pm$0.04   &                  &&                  &    151$\pm$3      &                   \\
00097564168 &  2025-01-18  &  08:58:38   &    172     &                   &                  & 11.15$\pm$0.04   &&                  &                   &   139$\pm$5       \\
00097564168 &  2025-01-18  &  09:03:12   &    338     &  11.08$\pm$0.04   &                  &                  &&    197$\pm$6     &                   &                   \\
00097564174 &  2025-01-28  &  02:12:25   &    259     &                   & 11.33$\pm$0.04   &                  &&                  &    137$\pm$3      &                   \\
00097564174 &  2025-01-28  &  02:16:53   &    172     &                   &                  & 11.15$\pm$0.04   &&                  &                   &   139$\pm$5       \\
00097564174 &  2025-01-28  &  02:21:27   &    323     &  11.18$\pm$0.04   &                  &                  &&    181$\pm$6     &                   &                   \\
00097564180 &  2025-02-07  &  20:59:42   &    274     &                   & 10.52$\pm$0.04   &                  &&                  &    287$\pm$7      &                   \\
00097564186 &  2025-02-17  &  15:51:59   &    276     &                   & 10.72$\pm$0.04   &                  &&                  &    240$\pm$5      &                   \\
00097564192 &  2025-02-27  &  18:03:28   &    248     &                   & 10.52$\pm$0.04   &                  &&                  &    287$\pm$7      &                   \\
00097564198 &  2025-03-09  &  03:36:25   &    275     &                   & 10.92$\pm$0.04   &                  &&                  &    199$\pm$4      &                   \\
00097564204 &  2025-03-19  &  16:39:18   &    262     &                   & 11.27$\pm$0.04   &                  &&                  &    145$\pm$3      &                   \\
00097564204 &  2025-03-19  &  16:43:49   &    174     &                   &                  & 11.08$\pm$0.04   &&                  &                   &   148$\pm$6       \\
00097564204 &  2025-03-19  &  16:48:26   &    325     &  11.11$\pm$0.04   &                  &                  &&    192$\pm$6     &                   &                   \\
00097564210 &  2025-03-29  &  20:42:06   &    277     &                   & 11.16$\pm$0.04   &                  &&                  &    159$\pm$4      &                   \\
00098285001 &  2025-04-03  &  14:43:09   &    749     &  11.03$\pm$0.04   &                  &                  &&    207$\pm$6     &                   &                   \\
00098285001 &  2025-04-03  &  14:48:38   &    437     &                   & 11.18$\pm$0.04   &                  &&                  &    156$\pm$3      &                   \\
00098285004 &  2025-04-18  &  13:45:36   &    448     &                   &                  & 11.02$\pm$0.04   &&                  &                   &   156$\pm$6       \\
00098285004 &  2025-04-18  &  13:55:06   &    897     &  11.03$\pm$0.04   &                  &                  &&    207$\pm$6     &                   &                   \\
00098285004 &  2025-04-18  &  14:06:51   &    684     &                   & 11.23$\pm$0.04   &                  &&                  &    150$\pm$3      &                   \\
00098285007 &  2025-05-03  &  20:27:19   &    444     &                   &                  & 11.34$\pm$0.04   &&                  &                   &   116$\pm$4       \\
00098285007 &  2025-05-03  &  20:36:37   &    890     &  11.57$\pm$0.04   &                  &                  &&    126$\pm$4     &                   &                   \\
00098285007 &  2025-05-03  &  20:48:04   &    544     &                   & 11.47$\pm$0.04   &                  &&                  &    120$\pm$3      &                   \\

\hline
\end{longtable}

\clearpage
\newpage
\onecolumn
\section{Atlas of high-resolution emission line profiles of T~CrB during SAP}
\label{appx:atlasSAP}

In this appendix we present an atlas of high-resolution profiles of selected
emission lines (H$\alpha$, H$\beta$, HeII 4686, HeI 5876, HeI 6678, and [OIII]
5007), obtained with the Asiago 1.82m + Echelle and CTIO 1.55m + Chiron
telescopes, covering the SAP phase of T~CrB from April 2015 to June 2023.
All spectra have been continuum normalized and are plotted on the same
ordinate scale for an easier comparison. The orbital phase quoted for each
spectrum has been computed according to the to the orbital solution given 
in Table~\ref{tab:orbit}, where phase 0.0 corresponds to the passage of the
RG at the ascending quadrature (maximum RG velocity).

   \begin{figure*}[h!]
   \centering
   \includegraphics[angle=270,width=14cm]{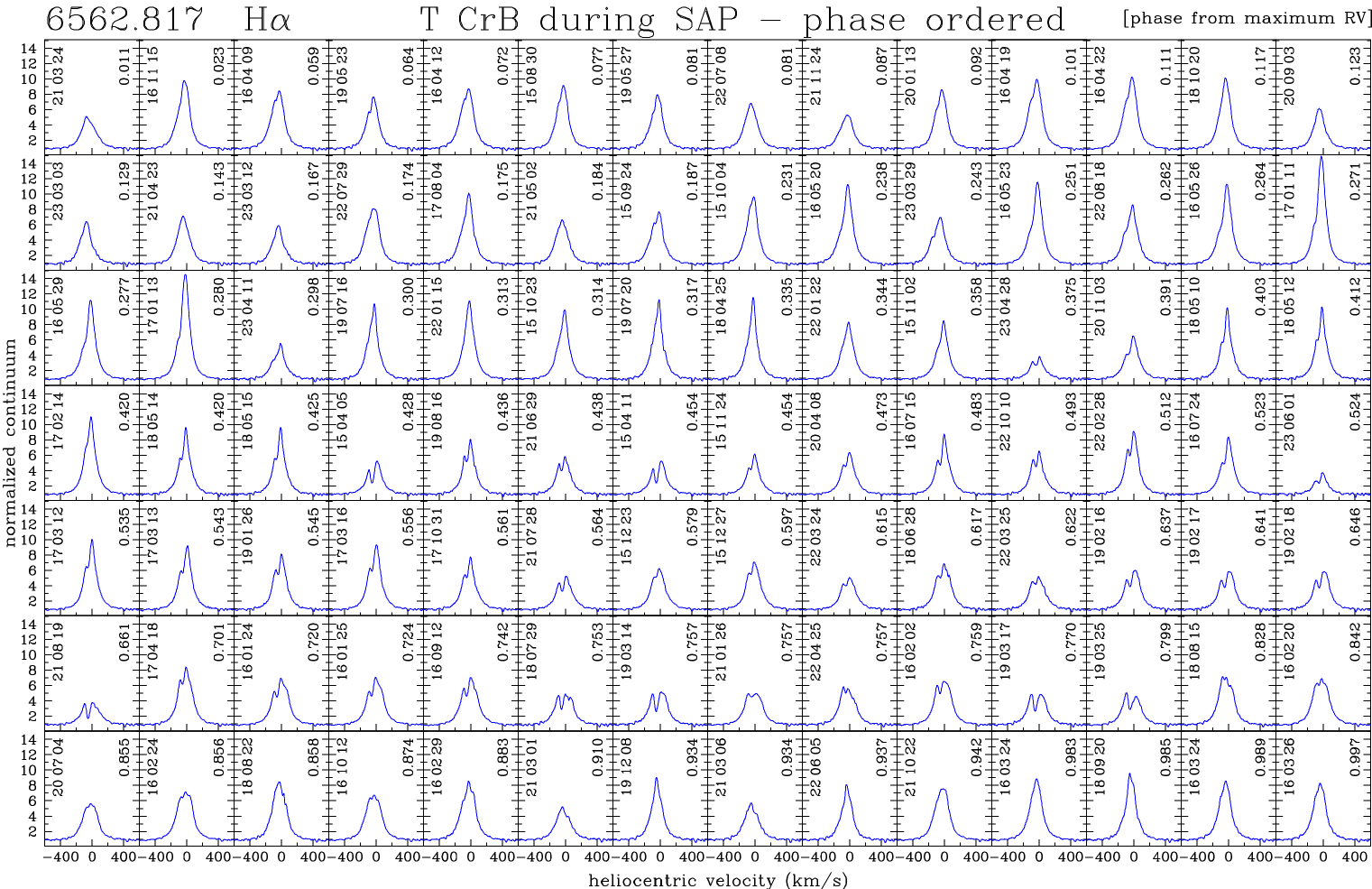}
      \caption{High-resolution H$\alpha$ emission line profiles of T~CrB during SAP
               from Asiago 1.82m + Echelle and SMARTS 1.55 + Chiron observations.}
         \label{fig:atlasHa}
   \end{figure*}

   \begin{figure*}[h!]
   \centering
   \includegraphics[angle=270,width=14cm]{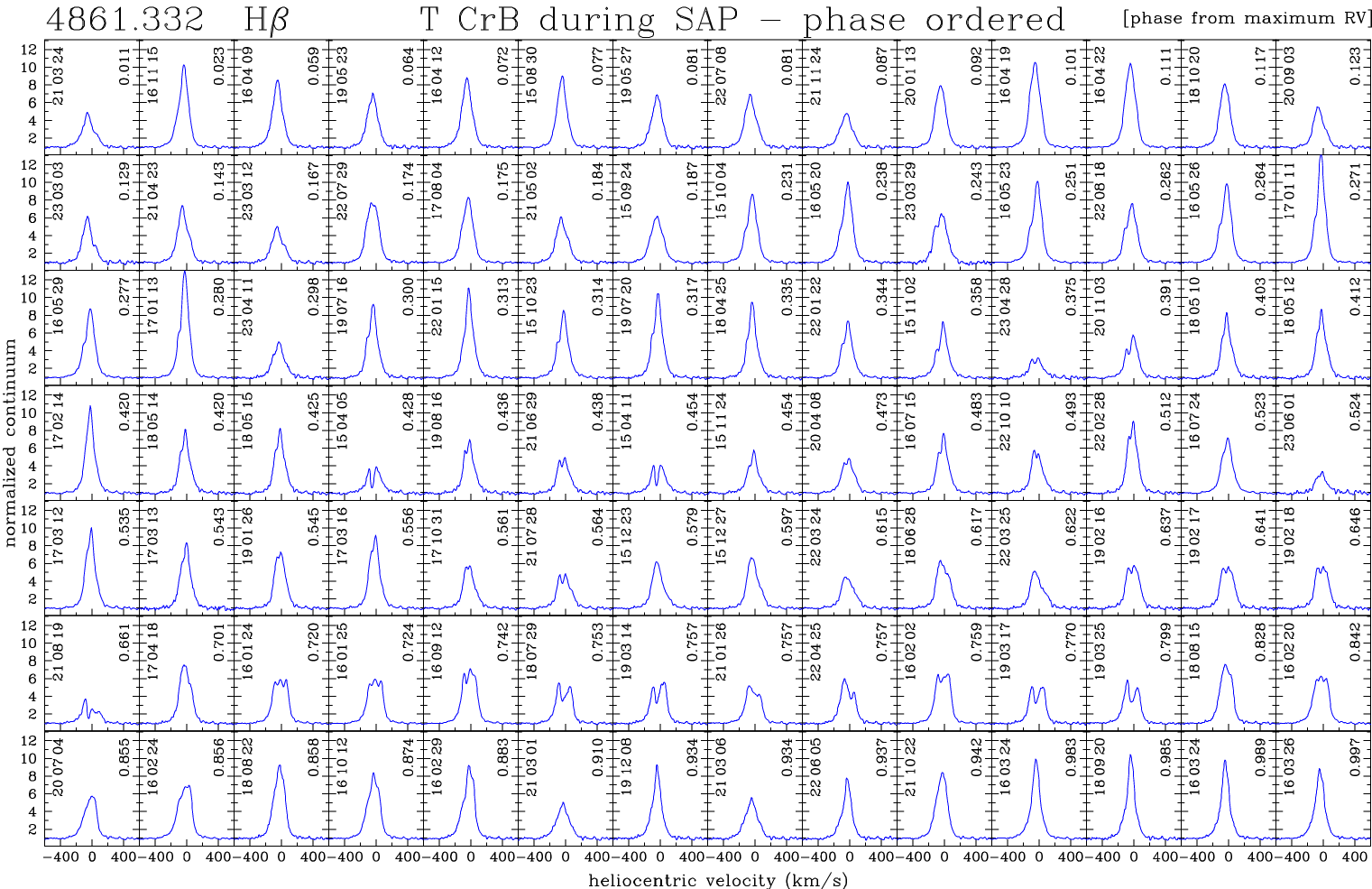}
      \caption{High-resolution H$\beta$ emission line profiles of T~CrB during SAP
               from Asiago 1.82m + Echelle and SMARTS 1.55 + Chiron observations.}
         \label{fig:atlasHb}
   \end{figure*}

   \begin{figure*}[h!]
   \centering
   \includegraphics[angle=270,width=14cm]{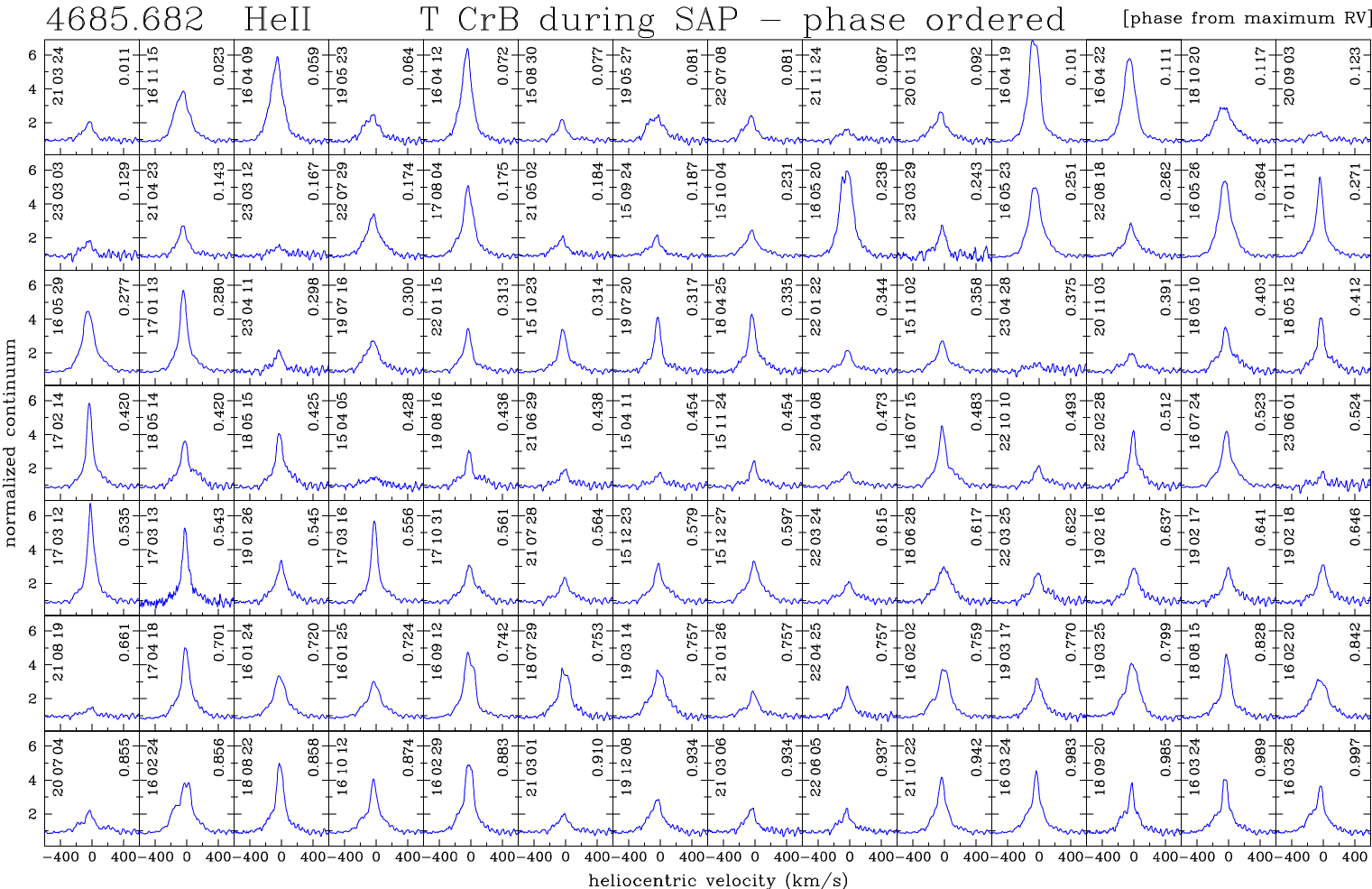}
      \caption{High-resolution HeII 4686\AA\ emission line profiles of T~CrB during SAP
               from Asiago 1.82m + Echelle and SMARTS 1.55 + Chiron observations.}
         \label{fig:atlasHeII}
   \end{figure*}

   \begin{figure*}[h!]
   \centering
   \includegraphics[angle=270,width=14cm]{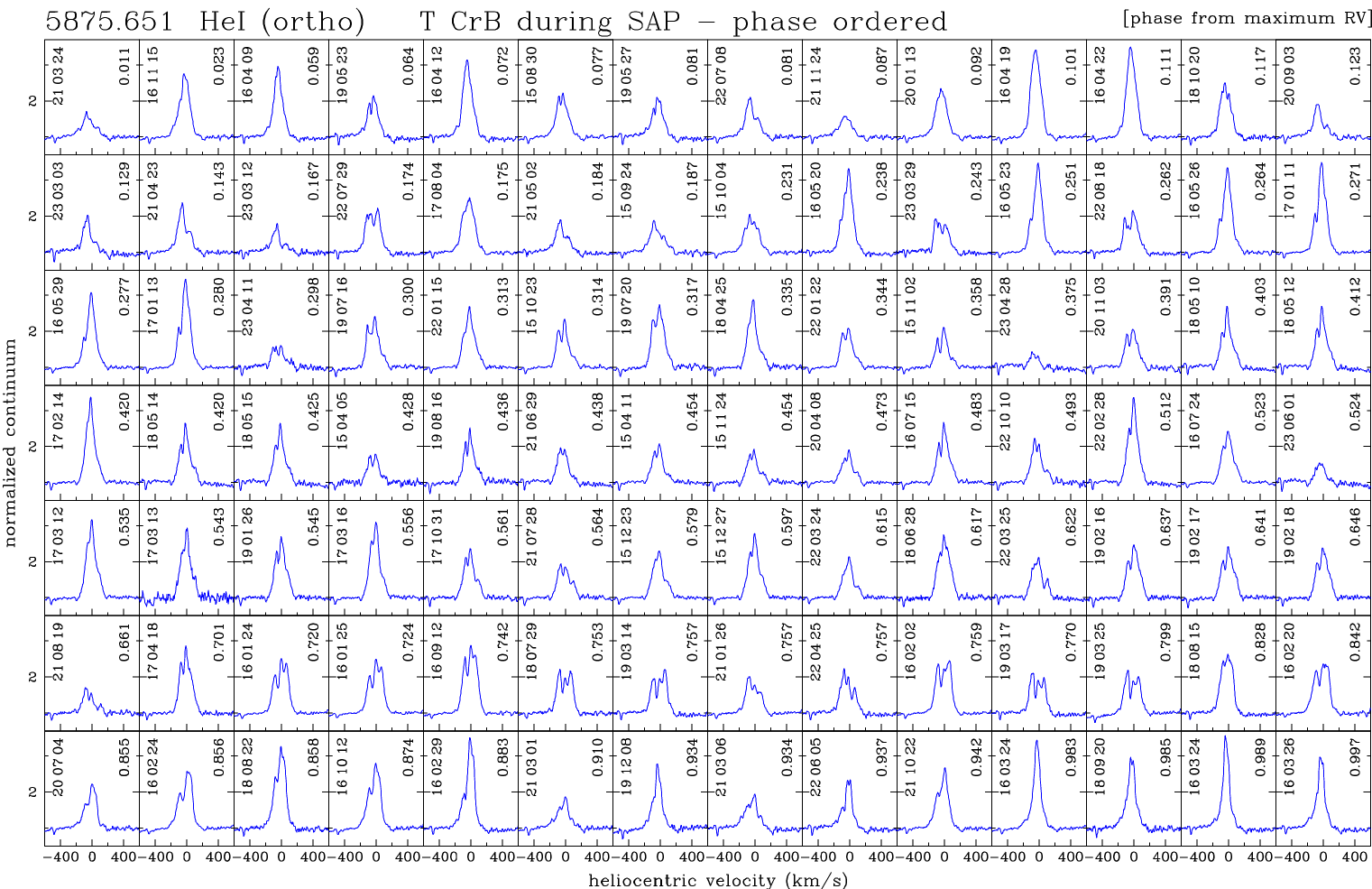}
      \caption{High-resolution HeI 5876\AA\ emission line profiles of T~CrB during SAP
               from Asiago 1.82m + Echelle and SMARTS 1.55 + Chiron observations.}
         \label{fig:atlas5876}
   \end{figure*}

   \begin{figure*}[h!]
   \centering
   \includegraphics[angle=270,width=14cm]{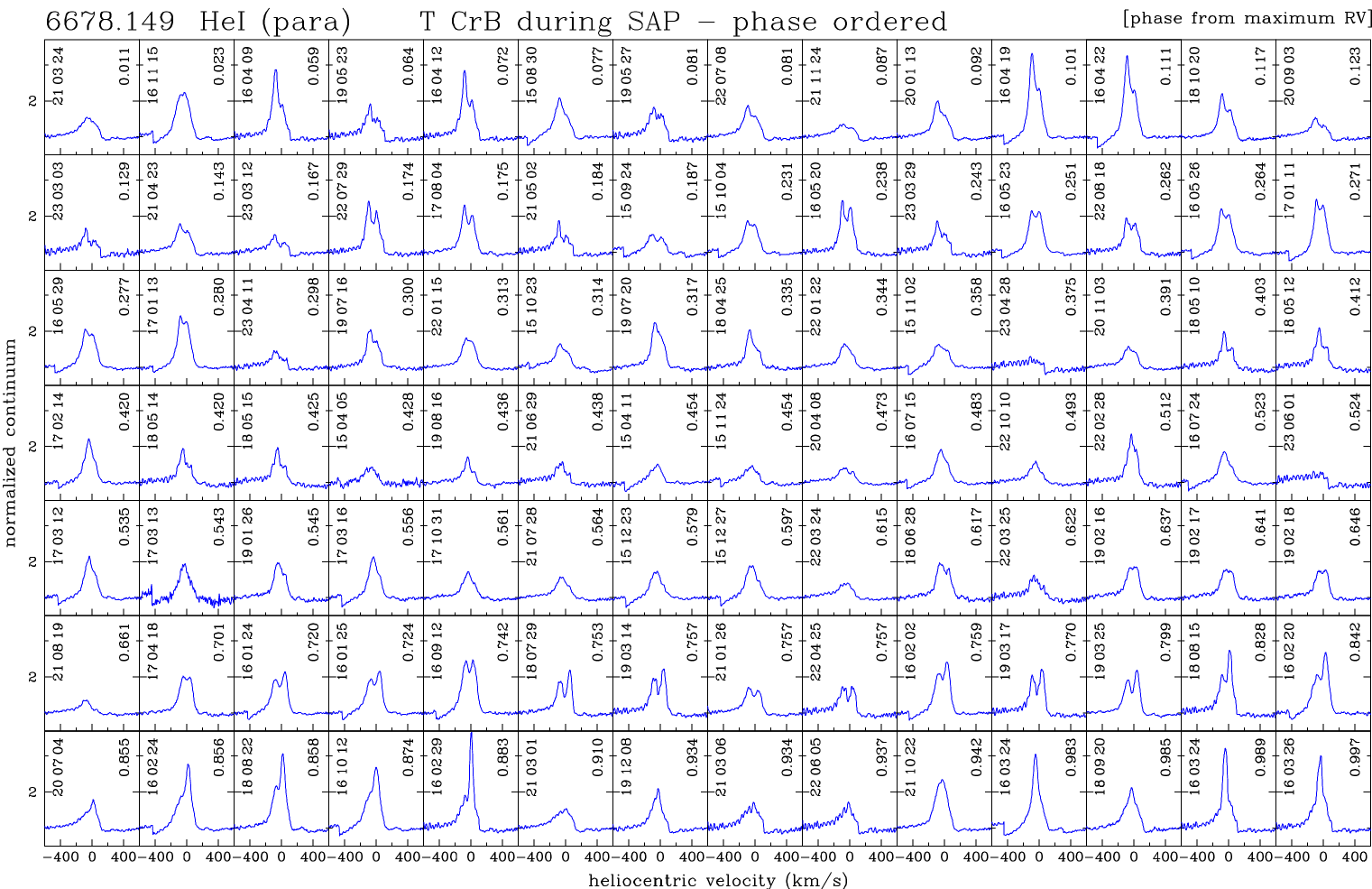}
      \caption{High-resolution HeI 6768\AA\ emission line profiles of T~CrB during SAP
               from Asiago 1.82m + Echelle and SMARTS 1.55 + Chiron observations.
               The occasional dent on the blue side of the line is caused by
               imperfect conjunction of adjacent Echelle orders.}
         \label{fig:atlas6678}
   \end{figure*}

   \begin{figure*}[h!]
   \centering
   \includegraphics[angle=270,width=14cm]{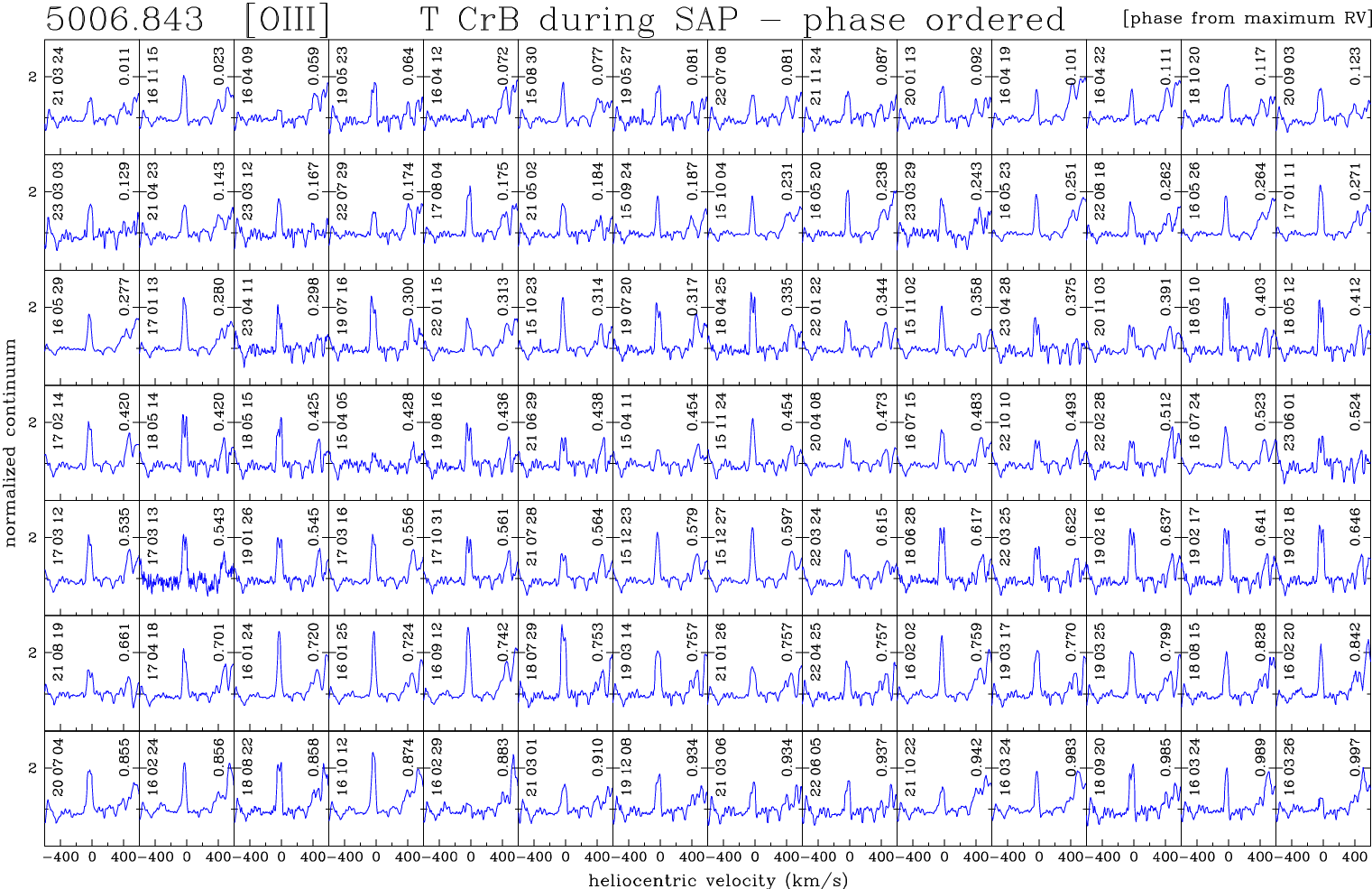}
      \caption{High-resolution [OIII] 5007\AA\ emission line profiles of T~CrB during SAP
               from Asiago 1.82m + Echelle and SMARTS 1.55 + Chiron observations.}
         \label{fig:atlas5007}
   \end{figure*}

\clearpage
\newpage
\onecolumn
\section{Atlas of high-resolution profiles for H$\alpha$ of T~CrB during
post-SAP and recovery phases}
\label{appx:atlaspost}

In this appendix we present a selection of high-resolution profiles for the
H$\alpha$ emission line of T~CrB, obtained with the Varese 0.84m telescope +
Multi-Mode Echelle spectrograph at a resolving power of 12,000.  These
profiles cover the post-SAP evolution of H$\alpha$ to present time, and thus extend the
time coverage of Fig.~\ref{fig:atlasHa} that focuses on the SAP phase alone.

To derive clean profiles, for all the observations presented in this picture
we have subtracted from the original spectra the spectrum of a template
M3III giant obtained with the same telescope, shifted to the epoch velocity
of the red giant in T~CrB.  The template spectrum is shown for reference at
the bottom of the panels.  Such a subtraction is particoularly important to
reconstruct the true profile at the time of the weakest emission in
H$\alpha$, during the deep-minimum phase discussed in
sect.~\ref{deepminimum}.

The spectra, distributed over three orbits, are aligned on the three panels
at the same orbital phase for ease of comparison (phase 0.0 corresponding to
passage of the red giant at ascending quadrature, as per
Table~\ref{tab:orbit}).  

The upper right box illustrates (at a more dense time mapping) the rapid
spectral H$\alpha$ evolution during the November 2024 flare (cf.
sect.~\ref{recovery}).  The same flare is well
visible also in the evolution of all other emission lines and particularly
in HeII 4686.  For sake of completeness, the bottom left box shows some spectra
recorded with Asiago and CTIO telescopes in the first six months of the post-SAP 
period, before the monitoring with the Varese 0.84m telescope could begin.

   \begin{figure*}[h!]
   \centering
   \includegraphics[width=\hsize]{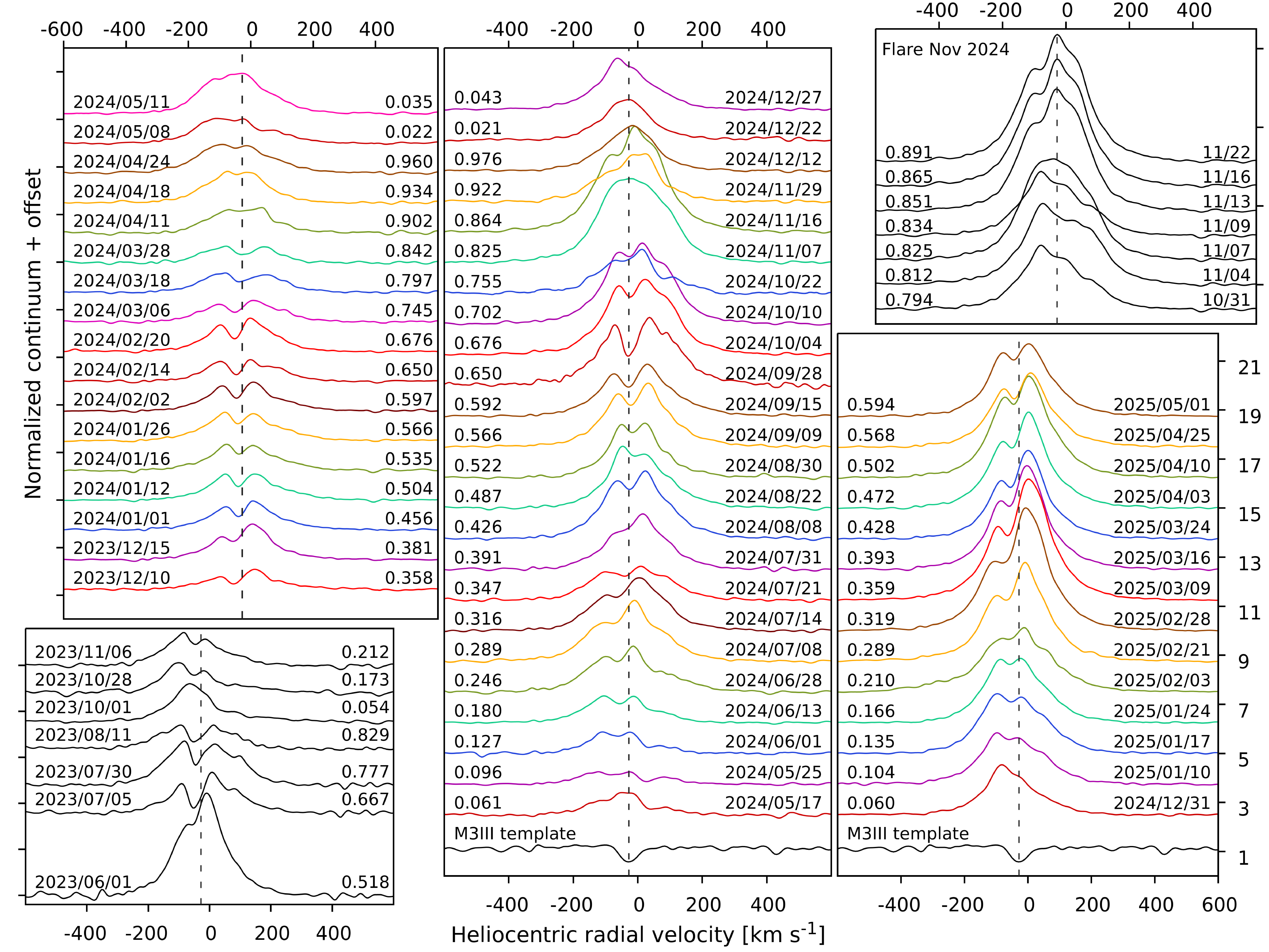}
      \caption{Evolution of the profile of H$\alpha$ emission line during the
               post-SAP phase of T~CrB, from Echelle observations collected with the
               Varese 0.84m telescope. The dashed vertical line marks the
               barycentric velocity of the system.}
         \label{fig:Uemflux}
   \end{figure*}

\clearpage
\newpage
\onecolumn
\section{Radial velocity measurements of the cool giant in T~CrB}
We report here the heliocentric radial velocity (RV) of the RG that has been measured 
via cross-correlation on the high-resolution spectra of T~CrB that we collected with the Asiago
1.82m + Echelle and SMARTS 1.5m + CHIRON telescopes.  The
template for cross-correlation was selected from the synthetic spectral
library of \citet{2005A&A...442.1127M} for the resolving power 20,000, with
atmospheric parameters $T_{\rm eff}$=3500~K, $\log g$=1.5, [Fe/H]=0.0,
[$\alpha$/H]=0.0, $\zeta$=2~km\,s$^{-1}$.
\begin{table}[h!]
\caption{\label{tab:rv}Radial velocities of the cool giant in T~CrB}
\label{tab:appendix-RV}
\centering
\begin{tabular}{crcr|ccrc}
\hline\hline
\multicolumn{4}{c|}{Asiago 1.82m+ Echelle}&\multicolumn{4}{c}{SMARTS 1.5m + CHIRON}\\
HJD & RV & HJD & RV && HJD & RV &\\
($-$2400000) & (km/s) &($-$2400000) & (km/s) &&($-$2400000) & (km/s) &\\
\hline
 55665.411 &   $-$4.92  & 57862.396 &  $-$32.65    & & 57448.887 &   $-$9.16 &  \\ 
 55989.662 &  $-$51.67  & 57970.333 &  $-$16.09    & & 57472.912 &   $-$2.76 &  \\ 
 56057.361 &  $-$23.91  & 58058.241 &  $-$50.96    & & 57474.815 &   $-$3.36 &  \\ 
 57118.610 &  $-$51.13  & 58234.355 &  $-$41.36    & & 57488.760 &   $-$5.19 &  \\ 
 57122.523 &  $-$50.56  & 58298.389 &  $-$47.43    & & 57491.781 &   $-$5.97 &  \\ 
 57265.305 &   $-$7.40  & 58329.370 &  $-$26.73    & & 57529.660 &  $-$25.76 &  \\ 
 57290.347 &  $-$17.96  & 58353.362 &  $-$11.36    & & 58249.709 &  $-$47.62 &  \\ 
 57300.283 &  $-$25.11  & 58382.267 &   $-$2.15    & & 58251.719 &  $-$48.43 &  \\ 
 57319.240 &  $-$39.25  & 58412.235 &   $-$6.87    & & 58253.708 &  $-$49.36 &  \\ 
 57329.222 &  $-$43.63  & 58509.684 &  $-$49.20    & & 58254.705 &  $-$49.57 &  \\ 
 57351.190 &  $-$49.53  & 58530.583 &  $-$42.51    & & 58346.491 &  $-$16.81 &  \\ 
 57351.195 &  $-$49.49  & 58531.549 &  $-$43.97    & & 58557.842 &  $-$27.69 &  \\ 
 57386.700 &  $-$47.65  & 58532.572 &  $-$42.88    & & 58560.883 &  $-$25.72 &  \\ 
 57411.655 &  $-$31.07  & 58566.517 &  $-$18.74    & & 58627.725 &   $-$6.20 &  \\ 
 57411.726 &  $-$32.90  & 58685.433 &  $-$38.15    & & 58631.683 &   $-$7.34 &  \\ 
 57412.685 &  $-$30.10  & 58712.319 &  $-$50.22    & & 59280.882 &   $-$5.77 &  \\ 
 57412.701 &  $-$30.34  & 58825.729 &   $-$6.34    & & 59337.742 &  $-$21.07 &  \\ 
 57419.642 &  $-$27.54  & 58863.723 &   $-$8.81    & & 59395.570 &  $-$50.00 &  \\ 
 57439.487 &  $-$12.68  & 58948.465 &  $-$51.85    & & 59639.904 &  $-$50.76 &  \\ 
 57442.679 &  $-$12.07  & 59035.360 &  $-$13.31    & & 59664.882 &  $-$44.35 &  \\ 
 57471.615 &   $-$3.12  & 59096.289 &  $-$13.40    & & 59695.756 &  $-$26.73 &  \\ 
 57498.394 &  $-$11.76  & 59157.245 &  $-$46.70    & & 59736.659 &   $-$5.91 &  \\ 
 57500.539 &   $-$7.77  & 59240.683 &  $-$24.26    & & 59790.486 &  $-$17.19 &  \\ 
 57534.469 &  $-$30.41  & 59298.456 &   $-$1.48    & & 59808.489 &  $-$28.84 &  \\ 
 57534.505 &  $-$32.05  & 59328.367 &  $-$12.99    & & 59810.486 &  $-$29.79 &  \\ 
 57535.352 &  $-$30.49  & 59424.326 &  $-$51.02    & & 60007.896 &  $-$10.06 &  \\ 
 57537.477 &  $-$32.03  & 59446.310 &  $-$41.53    & & 60033.800 &  $-$25.26 &  \\ 
 57585.365 &  $-$51.80  & 59510.231 &   $-$6.32    & & 60063.771 &  $-$42.93 &  \\ 
 57594.355 &  $-$52.49  & 59543.188 &   $-$5.02    & & 60097.671 &  $-$52.08 &  \\ 
 57644.278 &  $-$28.06  & 59594.737 &  $-$36.43    & &           &           &  \\ 
 57674.236 &  $-$14.38  & 59601.700 &  $-$39.93    & &           &           &  \\ 
 57708.202 &   $-$6.67  & 59601.709 &  $-$39.94    & &           &           &  \\ 
 57761.729 &  $-$29.73  & 59663.474 &  $-$45.25    & &           &           &  \\ 
 57762.709 &  $-$29.42  & 59769.354 &   $-$7.97    & &           &           &  \\ 
 57797.607 &  $-$50.57  & 59863.267 &  $-$49.71    & &           &           &  \\ 
 57822.580 &  $-$50.98  & 60016.468 &  $-$15.10    & &           &           &  \\ 
 57829.428 &  $-$52.34  & 60046.411 &  $-$33.48    & &           &           &  \\ 
\hline
\end{tabular}
\end{table}

\end{appendix}
\end{document}